\documentclass[twocolumn,           % Format : preprint, twocolumn
               showpacs,            % Pacs : showpacs, noshowpacs
               nopreprintnumbers,     % Preprint: preprintnumbers,
               aps,                 % Society: ...
               prd,          	    % Journal Style : pra, prb, prc, prd, pre,
               letterpaper,             % Size : a4paper, ...
               nofootinbib,         % Footnote: footinbib, nofootinbib
               tightenlines,        % Remove additional spaces in a line
               floats,floatfix,      % Floating pictures and tables
               showkeys
               ]{revtex4-1}

\usepackage{graphicx}% Include figure files
\usepackage{dcolumn}% Align table columns on decimal point
\usepackage{bm}% bold math
\usepackage{amsmath}
\usepackage{amsfonts,amssymb}
\usepackage{color}
\usepackage{enumerate}
\usepackage{booktabs}

\usepackage{float} % para usar[H]
\newcommand{\ra}[1]{\renewcommand{\arraystretch}{#1}}

\begin{document}

\title{Energy density profile inspired by noncommutativity}

\author{Miguel A. Garc\'{\i}a-Aspeitia$^{1,2}$} 
\email{aspeitia@fisica.uaz.edu.mx}

\author{J. C. L\'opez-Dom\'inguez$^{2}$}
%\email{jlopez@fisica.uaz.edu.mx}

\author{C. Ortiz$^{2}$}
%\email{ortizgca@fisica.uaz.edu.mx}

\author{Sinhue Hinojosa-Ruiz$^{2}$}
%\email{sinhue@fisica.uaz.edu.mx}

\affiliation{$^{1}$Consejo Nacional de Ciencia y Tecnolog\'ia, Av. Insurgentes Sur 1582. Colonia Cr\'edito Constructor, Del. Benito Ju\'arez C.P. 03940, M\'exico D.F. M\'exico}
\affiliation{$^{2}$Unidad Acad\'emica de F\'isica, Universidad Aut\'onoma de Zacatecas, Calzada Solidaridad esquina con Paseo a la Bufa S/N C.P. 98060, Zacatecas, M\'exico.}

\author{Mario A. Rodr\'iguez-Meza$^{3}$}
%\email{marioalberto.rodriguez@inin.gob.mx}

\affiliation{$^{3}$Departamento de F\'\i sica, Instituto Nacional de Investigaciones Nucleares, Apdo. Postal 18-1027, M\'exico D.F. 11801, M\'exico.}

\begin{abstract}
An important consequence which comes from noncommutativity (NC) is undoubtedly the energy density characterized by a microscopic free parameter; indeed a Trans-Planckian parameter. 
However, its functional form is an interesting and useful equation which can be analyzed in  astrophysical scenarios giving now astrophysical constraints. In this sense, this paper is devoted to explore the astrophysical consequences of an energy density with the same functional form of NC; mainly in stellar dynamics and rotation curves of galaxies. We start exploring toy models of stars with incompressible and polytropic fluids respectively, with the addition and coexistence with this new fluid. In both cases, we propose an appropriate constriction based on the difference between a correct and an anomalous behavior.
As a complement, we explore the rotation curves of galaxies assuming that the halo is a fluid with the same characteristic of a NC equation, obtaining the range of values for the free parameter through the analysis of eighteen LSB galaxies. Our results are compared with traditional models studied in literature like Pseudoisothermal (PISO), Navarro-Frenk-White (NFW), Burkert and WaveDM dark matter models. Finally, we have computed the surface density, $\rho_i r_i$ for each dark matter model, where $i$ is for PISO, NFW, Burkert, WaveDM and NC macroscopic version. In the later case, following the results found using SPARC galaxy catalog, we have found a theoretical value of $116.97 M_\odot$ pc$^{-2}$ while the data analysis gives us a value of $144.21 M_\odot$ pc$^{-2}$. 
The values of the surface density $\rho_i r_i$ are roughly constant and their mean values depend on the dark matter model.
Also we have computed the mass of each dark matter model within $300$ pc and found that there is a common mass for spiral galaxies of the order of $10^7 M_\odot$, that is in agreement with the results for dSph Milky Way satellites. 
This would give a central density for the halo of
$\sim 0.1 M_\odot$ pc$^{-2}$ 
 independent of the dark matter model.
\end{abstract}

%\draft
\date{\today}
\maketitle

%%%%%%%%%%%%%%%%%%%
\section{Introduction}	
%%%%%%%%%%%%%%%%%%%

One of the most successful outcomes of string theory is that the space-time coordinates may become NC operators and lead us
to a NC version of a gauge theory via the Seiberg and Witten map \cite{Seiberg}. After this result was published, a large number of papers related with NC in gauge theories, quantum mechanics, classical mechanics, quantum field theory and gravity has been published. In addition it is possible to note that the NC parameter has been bounded by different observations or experiments, mainly leading to an extremely small value \cite{Romero}. 

A different NC formulation of quantum field theory, based on coherent state formulation, can be achieved using the Feynman path integral on the NC plane which is a used framework for quantum mechanics and field theory \cite{Nicolini:2005vd}. In a recent paper \cite{Smailagic:2003yb} the authors discuss the gravitational analogue of the NC modification of quantum field theory, pointing out that NC is an intrinsic property of the manifold itself and affects gravity in an indirect way. The energy-momentum density determines space-time curvature. Thus in General Relativity (GR), the effects of NC can be taken into account by keeping the standard form of Einstein curvature tensor and introducing a modified energy-momentum tensor. The NC eliminates the point-like structures and replace them by smeared objects. The effect of smearing is implemented by using a Gaussian distribution of minimal width $\sqrt{\gamma}$ instead of a Dirac-delta function where the spherically symmetric and particle-like gravitational source can be written as:

\begin{equation}
\rho_\gamma(r)=\frac{M_{NC}}{(4\pi\gamma)^{3/2}}\exp\left(-\frac{r^2}{4\gamma}\right), \label{0}
\end{equation}
where $M_{NC}$ is the total mass parameter and $\gamma$ the NC parameter of the energy density related with the smearing of the NC energy density distribution. In traditional literature, NC free parameter has only been studied in quantum scenarios, constraining the value at Trans-Planckian scales \cite{Alboteanu:2006hh,MohammadiNajafabadi:2006iu,He:2006yy}, only having important effects in quantum gravity theories.

Hence, in order to study large scales, we relate the following functions and we change the NC parameter which comes from a quantum theory for a new macroscopic parameter $\theta$, applicable to astrophysical issues,  following the recipe:
\begin{equation}
\rho_\gamma(r)\to\rho_\theta(r), \;\; M_{NC}\to M_{NED} \;\; \rm and \;\; \gamma\to\theta. \label{1}
\end{equation}
The parameter are constrained by observations or under arguments of a behavior that fits with the the traditional literature. Indeed, if we assume that this fluid permeates the galaxies structures, it is natural to study the presence of this fluid, also in the stellar dynamics.

Therefore, to start the study we will call this new fluid as noncommutativity energy density (NED) for the inspiration that comes from NC; then we give ourselves to the task of study the previous statements, proposing two exercises to analyze the dynamics in a scenario where baryons and NED coexist (without mutual interaction), i.e. $T_{\mu\nu}^{eff}=T_{\mu\nu}^{bar}+T_{\mu\nu}^{NED}$, where the first one is related with the baryonic matter and the second one is related with NED, producing the astrophysical dynamics observed and with this, constraint the NED parameter with observables. These two exercises analyze the new dynamic produced by the presence of NED, mainly in the evolution and dynamic of the stars and in rotation curves of galaxies.

In this sense, there has been extensive studies in stellar dynamics since the advent of GR \cite{weinberg:1972}; in such a way that observing the imprints of NED in the dynamics must be clear, showing deviations to GR predictions. For instance it is possible to assume that hypothetical NED particles are so heavy, forming a dense core in the star center allowing the applied treatment in the following sections. In this vein, stars with uniform density and white dwarfs are studied due that are excellent laboratories to study possible extensions to the GR background. It is important to remark that extensive studies about NC, can be seen in \cite{Hernandez-Almada:2016ioe}.

In addition, the functional form of Eq. \eqref{0} with the recipe \eqref{1}, inspire to be used as a density profile to reproduce the velocity rotation curves of galaxies, due to its similarity with Einasto's model \cite{Einasto} with $n^\prime=0.5$. In galactic dynamic the NED matter is concentrated mainly in the halo, with a Gaussian density profile and thus giving restrictions to the free parameter $\sqrt{\theta}$; constraining where NED parameter generates important effects. Other interesting dark matter (DM) proposals can be checked in Refs. \cite{piso,NFW,Burkert,Martins/Salucci:2007,Rodriguez-Meza:2012b,Guzman/Matos:2000} or for an excellent review see Ref. \cite{Rubin2001}.

Recapitulating, we remark that the recipe shown in Eq. \eqref{1} as a macroscopic phenomena, will generate constraints of the $\theta$ parameter also, of the order of macroscopic scales. This of course contributes to generate a macroscopic bounds, which was left out by Ref. \cite{Rahaman:2010vv}, contributing to strengthen these previous research.
In this sense, we will provide a table of the constraints of this parameter, which will be subject to the theoretical model under study or the observations with which it is contrasted.

From here, it is possible to organize the paper as follows: In Sec. \ref{SO}, we analyze a Newtonian star in two cases, where it is composed by an uniform density (incompressible fluid) coexisting with the NED matter and when it is composed by a polytropic matter and the core contains a NED fluid, showing a modified Lane-Emden equation, constraining the NED free parameter and fixing the NED mass as a subdominant component. In Sec. \ref{GC} we implement an analysis of galactic rotation curves, assuming that the DM halo can be modeled by NED density matter. In this case, we use a sample of eighteen LSB galaxies without photometry with the aim of constraint the NED parameters ($\rho_{NED}$ and $\sqrt\theta$) and compare with traditional density profile models like pseudoisothermal (PISO), Navarro-Frenk-White (NFW), Burkert and the WaveDM studied in these two Refs.\cite{2017arXiv170205103U,2017arXiv170100912B}, it is an ultralight scalar model motivated by large scale simulations\cite{2014NatPh..10..496S}. Finally in Sec. \ref{CR} we give a discussion and conclusions about the results obtained through the paper.

In what follows, we work in units in which $c=\hbar=1$, unless explicitly written.

%%%%%%%%%%%%%%%%%%%%%%%%%%%%%%%%
\section{Toy model stars with a NED component} \label{SO}
%%%%%%%%%%%%%%%%%%%%%%%%%%%%%%%% 

In this section, we study the stellar dynamics with a component of NED, together with the traditional baryonic matter. We start using the approach of a Newtonian star, composed by this fluid and baryonic matter with uniform density (incompressible fluid). After that, we study a most generic star through the Lane-Emden (LE) equation, under the assumption that the star is composed by a polytrope and also contains a NED. Both models are considered under the premise that ordinary matter does not interact with the NED and fixes one of the free parameters, with the argument of a subdominant NED.
 
%%%%%%%%%%%%%%%%%%%%%%%%%%%%%%%%
\subsection{NED matter on stars with uniform density in Newtonian approach}
%%%%%%%%%%%%%%%%%%%%%%%%%%%%%%%%

These stars are of interest, because they are simple enough to allow an exact solution in Newtonian background. Then, stars with uniform density consist of an incompressible fluid with equation of state (EoS), $\rho_I=\rm constant$, in such a way that a generalization with the addition of NED is simple, being excellent laboratories to understand the physics behind such processes.

In a Newtonian approach, dynamic equation for the evolution of a star can be written as:
\begin{equation}
r^2\frac{dp(r)}{dr}=-G\mathcal{M}(r)\rho_{eff}(r), \label{second}
\end{equation}
being $G$ the Newtonian gravitational constant, $\rho_{eff}(r)$ the effective density and $\mathcal{M}(r)$ the stellar mass written in the form:
\begin{equation}
\mathcal{M}(r)=4\pi \int_0^r r^{\prime2}\rho_{eff}(r^{\prime})dr^{\prime}. \label{first}
\end{equation}
Eqs. \eqref{second} and \eqref{first} are considered as the equations of motion for stellar dynamics in Newtonian approach. In addition, we propose that the star density is composed by an uniform component density and a NED in the following form:
\begin{equation}
\rho_{eff}(r)=\rho_{I}+\frac{M_{NED}}{(4\pi\theta)^{3/2}}\exp\left(-\frac{r^2}{4\theta}\right),
\end{equation}
where $\rho_I$ is a constant and there is not mutual interaction between the components. In order to have a numerical solution of the equations of motion we assume the following dimensionless variables:
\begin{subequations}
\begin{eqnarray}
&&x=\sqrt{\frac{GM}{R}}\left(\frac{r}{R}\right), \;\;  \bar{\mathcal{M}}(r)=\sqrt{\frac{G^{3}M}{R^{3}}}\mathcal{M}(r), \;\; \\ && \bar{p}(r)=\frac{4\pi R^3 }{M}p(r), \;  \bar{\theta}=\frac{4GM}{R^3}\theta, \;\;  \bar{\rho}_{I}=\frac{4\pi R^{3}}{3M}\rho_{I}, \;\; \\ && \bar{M}^{NED} = 8\sqrt{\frac{G^3M}{\pi R^3}}M_{NED},  \label{eq:8.3} 
\end{eqnarray}
\end{subequations}
and integrating Eq. \eqref{first} we have:
\begin{eqnarray}
&&\bar{\mathcal{M}}(x) = \bar{\rho}_Ix^3+\nonumber\\&&\frac{\bar{M}^{NED}}{\bar{\theta}^{3/2}}\left[\sqrt{\pi} \left(\frac{\bar{\theta}}{4}\right)^{3/2}{\rm Erf}\left(\frac{x}{\sqrt{\bar{\theta}}}\right)-\frac{\bar{\theta}}{4}x\exp\left(-\frac{x^2}{\bar{\theta}}\right)\right]\, \label{dim01} \, , 
\end{eqnarray}
where $\rm Erf(x)$ is the error function defined as: ${\rm Erf}(x)=(2/\sqrt{\pi})\int_0^x\exp(-t^2)dt$.

\begin{figure}[htbp]
\centering
\begin{tabular}{cc}
\includegraphics[scale=0.42]{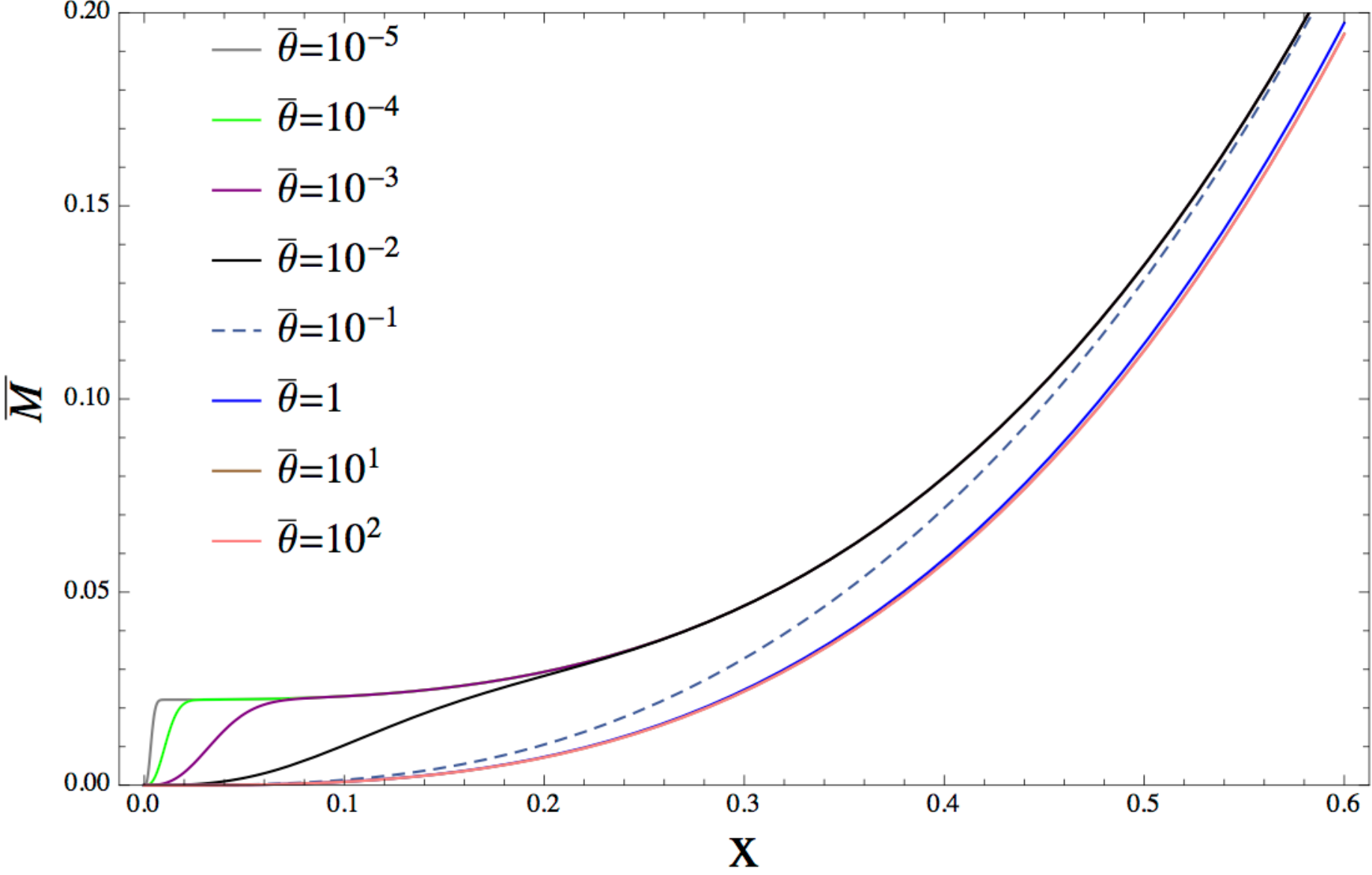} \\
\includegraphics[scale=0.38]{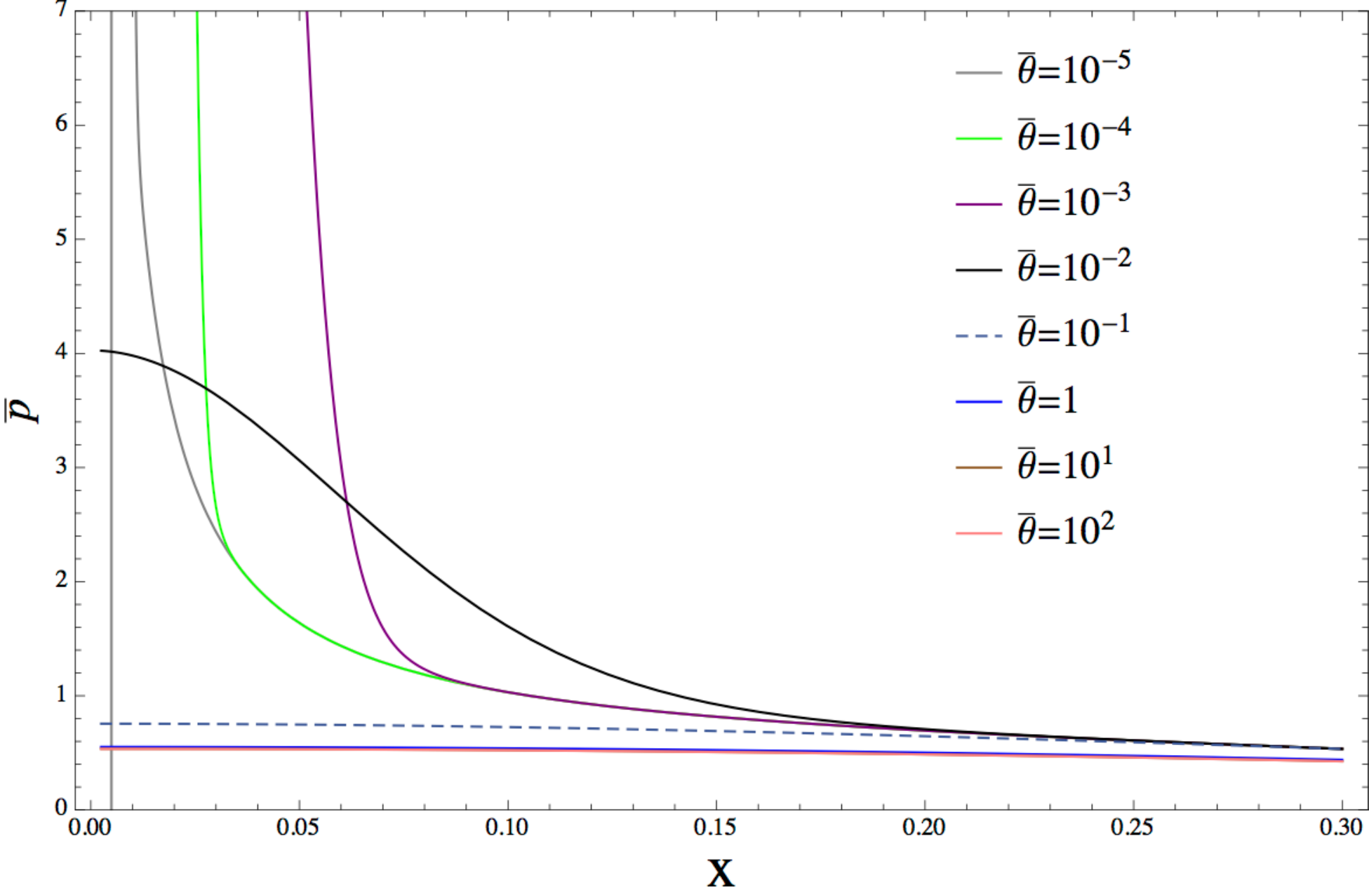}
\end{tabular}
\caption{Behavior of $\bar{\mathcal{M}}$ and $\bar{p}$ vs $x$, assuming the following free parameters: $\bar{\rho}_{B}=0.9$, $\bar{M}^{NED}=0.1$ and $GM/R=0.44$; together with the initial condition $\bar{p}(0.663)=0$ for Eq. \eqref{dim02}.  (Top) In this case it is possible to note how mass shows almost the traditional behavior when $\bar{\theta}<1$, other upper values, present important contributions that comes from NED. (Bottom) Pressure behavior for different values of $\bar{\theta}$, blue line is the threshold between the standard behavior and the behavior when NED play an important role for an incompressible fluid.} 
\label{fig1}
\end{figure}

The behavior can be seen in Fig. \ref{fig1} (Top) for different values of $\bar{\theta}$. 
In this case, we propose the extreme case for the free parameters in the form: $\bar{\rho}_I=0.9$ and $\bar{M}^{NED}=0.1$, considering domination of baryonic matter over NED component. It is important to mention that a greater amount of NED material, compromises the stability of the star.

Also, we analyze an extreme Newtonian star fulfilling the compactness relation $GM/R=0.44$. Indeed, it is possible to observe that the dimensionless NED parameter $\bar{\theta}$ dictates the behavior of the stellar mass.

In addition, Eq. \eqref{second} can be written in terms of dimensionless variables as:
\begin{eqnarray}
\frac{d\bar{p}(x)}{dx} = -\frac{\bar{\mathcal{M}}(x)}{x^2}\left[3\bar{\rho}_I+\frac{\bar{M}^{NED}}{\bar{2\theta}^{3/2}}\exp\left(-\frac{x^2}{\bar{\theta}}\right) \right], \label{dim02}
\end{eqnarray}
and its numerical integration can be observed in Fig. \ref{fig1} (Bottom) for different values of the NED parameter. It is notorious how for values below $\bar{\theta}=10^{-1}$, pressure and mass present important differences in comparison with the other cases when the NED parameter plays a role. 

Thus, small values of NED parameter shown the convergence to the traditional behavior without NED for pressure and mass in a Newtonian star with uniform density.

From both results we conclude that it is necessary that $\bar{\theta}<1$ to obtain the traditional behavior found in literature (stars with only uniform density) for pressure and mass. 
Then, based on the results shown in Figs. \ref{1} and with the aim of mimic the well known stellar behavior, it is possible to establish a bound of the NED parameter as: 
\begin{equation}
\sqrt{\theta}<\frac{1}{2}\sqrt{\frac{R}{GM}}R,
\end{equation}
where for a star in the limit of stellar stability as it is our case, we have $\sqrt{\theta}<0.753R$; which it is highly dependent on the stellar radius and the value must be macroscopic in order that fits to stellar dynamics.

%%%%%%%%%%%%%%%%%%%%%%%%%%%%%%%%%%%%%%%%
\subsection{Dwarf stars with NED}
%%%%%%%%%%%%%%%%%%%%%%%%%%%%%%%%%%%%%%%%

A dwarf star is mainly composed by a polytropic matter with equation of state (EoS) $P=K\rho^{(n+1)/n}$, being $K$ the polytropic constant and $n$ the polytropic index. This kind of stars are the most studied in literature due that it is possible to model them by just specifying the values of $K$ and $n$, having in mind the appropriate limits of both stellar structures\footnote{Neutron stars should be modeled by the Tolman-Oppenheimer-Vokoff equations, i.e., the strong field limit.}.

\begin{figure}[htbp]
\centering
\begin{tabular}{cc}
\includegraphics[scale=0.3]{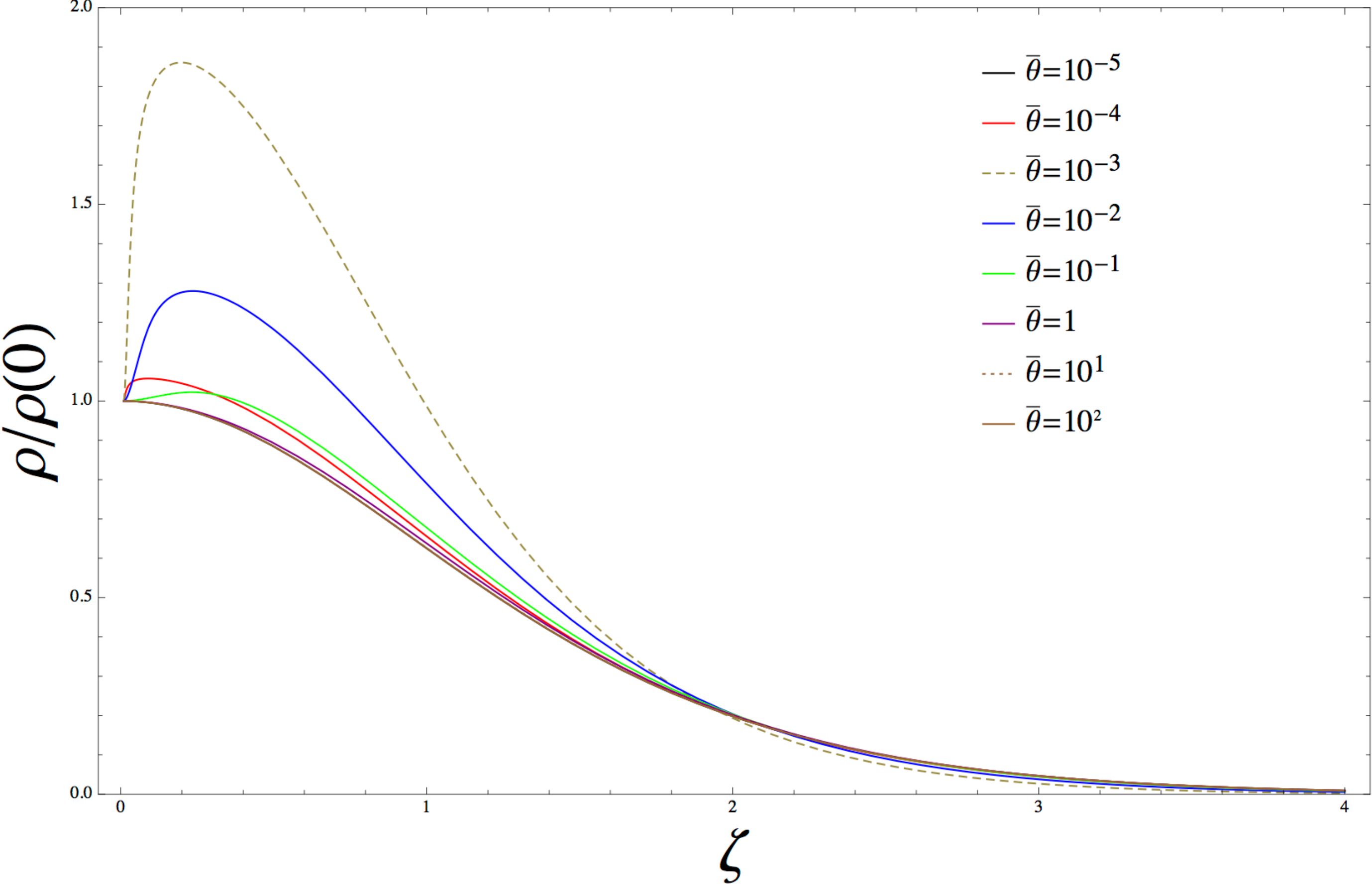} 
\end{tabular}
\caption{Behavior of $\Theta$ vs $\zeta$, with the initial conditions $\Theta(0)=1$, $\Theta^{\prime}(0)=0$. Top, from left to right and similarly from the figures in the bottom, we use $M^0_{NED}=0.1$.} 
\label{NCD}
\end{figure}

However now we consider a NED component in the effective density of the star. Starting from Eq. \eqref{second} it is possible to write the modified LE equation as:
\begin{equation}
\frac{1}{\zeta^2}\frac{d}{d\zeta}\zeta^2\frac{d\Theta}{d\zeta}+\Theta^n=\frac{M^0_{NED}}{\bar{\theta}^{3/2}}\exp\left(-(1+n)\frac{\zeta^2}{\bar{\theta}}\right),
\end{equation}
where it was proposed the following dimensionless variables:
\begin{subequations}
\begin{eqnarray}
 &&r=\zeta\left(\frac{K(1+n)}{4\pi G}\right)^{1/2}\rho(0)^{(1-n)/2n},  \rho=\rho(0)\Theta^n, \;\; \\ && P=K\rho(0)^{(1+n)/n}\Theta^{n+1}, \theta=\frac{K\rho(0)^{(1-n)/n}}{16\pi G}\bar{\theta}, \;\;  \\ &&M_{NED}=\left(\frac{4\pi K\rho(0)^{(1-n)/n}}{16\pi G}\right)^{3/2}M^0_{NED}. \label{eq:8.3} 
\end{eqnarray}
\end{subequations}
Notice that $\rho(0)$ is the density at the center of the star. Moreover, we assume the same initial conditions for the problem of LE as: $\Theta(0)=1$, $\Theta^{\prime}(0)=0$ and in case of Dwarf stars, it is assumed a polytropic index $n=3$. The results are shown in Fig. \ref{NCD}, restricting a subdominant $M^0_{NED}=0.1$ and $\bar{\theta}$ is related with the presence of NED term in LE equation and being the two free parameters of the theory. Here, the results are more restrictive, showing that for values $10^{-4}\geqslant\bar{\theta}$ and $\bar{\theta}\geqslant1$, it is recovered the traditional case. Therefore, using Eq. (11b), it is possible to constraint the NED parameter as
\begin{equation}
\kappa\geqslant10^2\sqrt{\theta} \;\; \rm and \;\;  \kappa\leqslant\sqrt{\theta},
\end{equation}
where
\begin{eqnarray}
\kappa\equiv\left[\frac{5}{192\pi^3 G}\left(\frac{3\pi^2}{m_N\mu}\right)^{4/3}\frac{1}{\rho(0)^{2/3}}\right]^{1/2},
\end{eqnarray}
being $\mu$ the number of nucleons per electron, $m_N$ is the nucleon mass \cite{weinberg:1972} and $\rho(0)$ is shown in Table \ref{tab:wdwarfs} for different Dwarf Stars reported in literature \cite{wd1,40erib,sirius}. Inside of the region, the NED presence is predominant and therefore, not having a traditional behavior of the stellar dynamics.

Also, we report the values of the NED parameters for the fifteen white dwarfs shown in Table \ref{tab:wdwarfs}, constraining the NED parameter with observables to obtain a traditional behavior shown in literature.

%%%%%%%%%%%%%%%%%%%%%%%%
% Sample of White Dwarfs
\begin{center}
\begin{table*}
\ra{1.3}
\begin{center}
\begin{tabular}{@{}l l r r r r r r r r r r r @ {}}\toprule
    \hline \hline
    \multicolumn{12}{c}{White Dwarf sample} \\
    \hline 
    \multicolumn{12}{c}{Observational data}\\
   \hline 
\multicolumn{1}{c}{White Dwarf} &
\multicolumn{1}{c}{Mass (M$_{\odot}$) }&
\multicolumn{1}{c}{Radius (R$_{\odot}$) } &
\multicolumn{1}{c}{$\rho(0)$ (MeV$^4$) } &
\multicolumn{1}{c}{$\sqrt{\theta}$($10^{18}$MeV$^{-1}$$\geqslant/\leqslant10^{-2}$)}& \\
 
\cmidrule[0.4pt](r{0.125em}){1-1}
\cmidrule[0.4pt](lr{0.125em}){2-2}
\cmidrule[0.4pt](lr{0.125em}){3-3}
\cmidrule[0.4pt](lr{0.125em}){4-4}
\cmidrule[0.4pt](lr{0.125em}){5-5}
\cmidrule[0.4pt](lr{0.125em}){6-6}
\cmidrule[0.4pt](lr{0.125em}){7-7}
\cmidrule[0.4pt](lr{0.125em}){8-8}
\cmidrule[0.4pt](lr{0.125em}){9-9}
\cmidrule[0.4pt](lr{0.125em}){10-10}
\cmidrule[0.4pt](lr{0.125em}){11-11}
\cmidrule[0.4pt](lr{0.125em}){12-12}
Sirius B & 1.034 & 0.0084 & 10.5993 & 3.87583 \\ 
Procyon B & 0.604 & 0.0096 & 4.1478 & 5.29888  \\ 
40 Eri B & 0.501 & 0.0136 & 1.21009 & 7.98947  \\ 
EG 50 & 0.50 & 0.0104 & 2.70063 & 6.11367  \\ 
GD 140 & 0.79 & 0.0085 & 7.81565 & 4.29011  \\ 
CD-38 10980 & 0.74 & 0.01245 & 2.3298 & 6.4222  \\ 
W485A &0.59 & 0.0150 & 1.06212 & 8.34448  \\ 
G154-B5B & 0.46 & 0.0129 & 1.3006 & 7.79966  \\ 
LP 347-6 & 0.56 & 0.0124 & 1.7827 & 7.02152 \\ 
G181-B5B & 0.54 & 0.0125 & 1.6781 & 7.16447  \\ 
WD1550+130 & 0.535 & 0.0211 & 0.3456 & 12.132 \\ 
Stein 2051B& 0.48 & 0.0111 & 2.13023 & 6.6168  \\ 
G107-70AB & 0.65 & 0.0127 & 1.926 & 6.84287 \\ 
L268-92 & 0.70 & 0.0149 & 1.28438 & 7.83236 \\ 
G156-64 & 0.59 & 0.0110 & 2.69047 & 6.12135 \\ 

 \bottomrule
\hline \hline
\end{tabular}
\end{center}
\caption{From left to right the columns read; name of the star, mass in solar units M$_{\odot}$, radius in R$_{\odot}$, density as $\rho(0) = 3M/4\pi R^3$ in ${\rm MeV}^4$ and NED parameter in ${\rm MeV}^{-1}$ deduced from the constraint mentioned in text. Here we use a catalogue of fifteen white dwarfs reported in \citep{wd1,40erib,sirius}.}\label{tab:wdwarfs}
\end{table*}
\end{center}
%%%%%%%%%%%%%%%%%%%%%%%%
%

%%%%%%%%%%%%%%%%%%%%%%%%%%%%%%%%%%%%%%%%
\section{NED in galaxy rotation velocities} \label{GC}
%%%%%%%%%%%%%%%%%%%%%%%%%%%%%%%%%%%%%%%%

To complement our analysis we study rotation curves of galaxies at the weak gravitational field limit in order to study  NED parameters. We start modeling the halo of the galaxy with the energy density of the NED shown in Eqs. \eqref{0} and \eqref{1}.

Moreover, it is important to notice some important characteristics of NED profile. 
Implementing an appropriate series expansion of the NED of the order $r\ll\sqrt{\theta}$, we generate a functional form similar to PISO \cite{piso}. 
In this context, this density profile is valid only at the center of galaxies. However, PISO density profile is an empirical profile designed for modeling DM in spirals galaxies and it has been applied not only at the center of galaxies but also in the outer spatial regions.
Also, it is possible to observe that NED is a particular case of Einasto's density profile \cite{Einasto} (see Eq. \eqref{Ein12}) when  $n^{\prime}=0.5$
\begin{equation}
\rho_{Ein}=\rho_{-2} \exp\left(-2n^{\prime}\left[ \left(\frac{r}{r_{-2}}\right)^{1/n^{\prime}}-1\right]\right), \label{Ein12}
\end{equation} 
where $r_{-2}$ is the radius where the density profile has a slope $-2$ and $\rho_{-2}$ is the local density at that radius; the parameter $n^{\prime}$ is known as Einasto index which describes the shape of the density profile. 

In general, the NED distribution can provide with extra information than other models can not (see for example \cite{piso,NFW,Burkert}), mainly due to the advantage that comes naturally from the geometric properties of space-time and it is not just chosen by observations or numerical simulations.

On the other hand, we have that the rotation velocity is obtained from the absolute value of the effective potential as:
\begin{equation}
V^2(r)=r\left\vert \frac{d\Phi(r)}{dr}\right\vert=\frac{G\mathcal{M}(r)}{r},  \label{rotvel}
\end{equation}
where $\Phi(r)$ is the gravitational potential and $\mathcal{M}(r)$ is the total mass which describes the galactic dynamics and it is expressed in the same way as it is shown in Eq. \eqref{first} or as given in Eq. \eqref{velDM} for the mass, $M_{DM}(r)$, steaming for a dark matter (DM) distribution.

%%%%%%%%%%%%%%%%%%%%%%%%%%%%%%%
\subsection{NED rotation velocity}
%%%%%%%%%%%%%%%%%%%%%%%%%%%%%%%

The rotation velocity for the NED can be obtained through the NED distribution, giving the following relationship:
\begin{eqnarray}
\hspace{-2em}
V^{2}_{\rm NED}(r) &=&
\frac{4\pi G\theta^{3/2}\rho_{NED}}{r} \nonumber \\
&& \times \left[ \sqrt{\pi}{\rm Erf}\left( \frac{r}{2\sqrt{\theta}}\right)- \frac{r}{\sqrt{\theta}}\exp\left(- \frac{r^2}{4\theta}\right)\right], \label{velrotnon}
\end{eqnarray}
where again, ${\rm Erf}(x)$ is the error function. This expression can be rewritten as
\begin{equation}
\hspace{-2em}
V^{2}_{\rm NED}(r) =
4\pi G \sqrt\theta\, \mu_{DM} \hat{V}(r/2\sqrt\theta) \label{velrotnonadim}
\end{equation}
where $\hat{V}(x)$ is a dimensionless function and we have defined $\mu_{DM}=\sqrt\theta\, \rho_{NED}$ (Eq. \eqref{muDM}),
which has turned out to be a very important quantity, characterizing the dark matter models in galaxies\cite{2009MNRAS.397.1169D}. It has units of surface density.
The rotation curve equations for PISO, NFW, Burkert and WaveDM models can also be written similarly as \eqref{velrotnonadim} (see Appendix \ref{Ap}).

\begin{figure}[htbp]
\centering
\begin{tabular}{cc}
\includegraphics[scale=0.85]{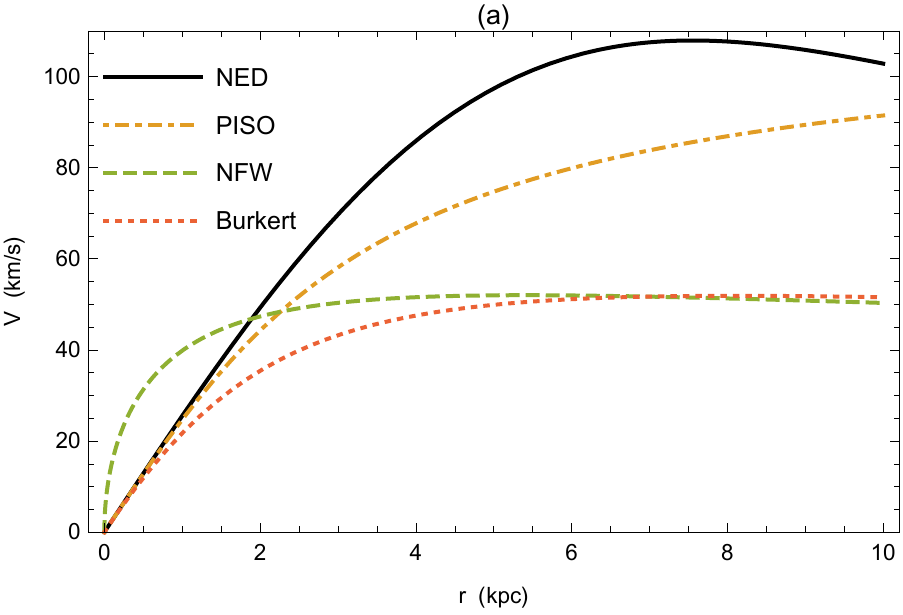} \\
\includegraphics[scale=0.85]{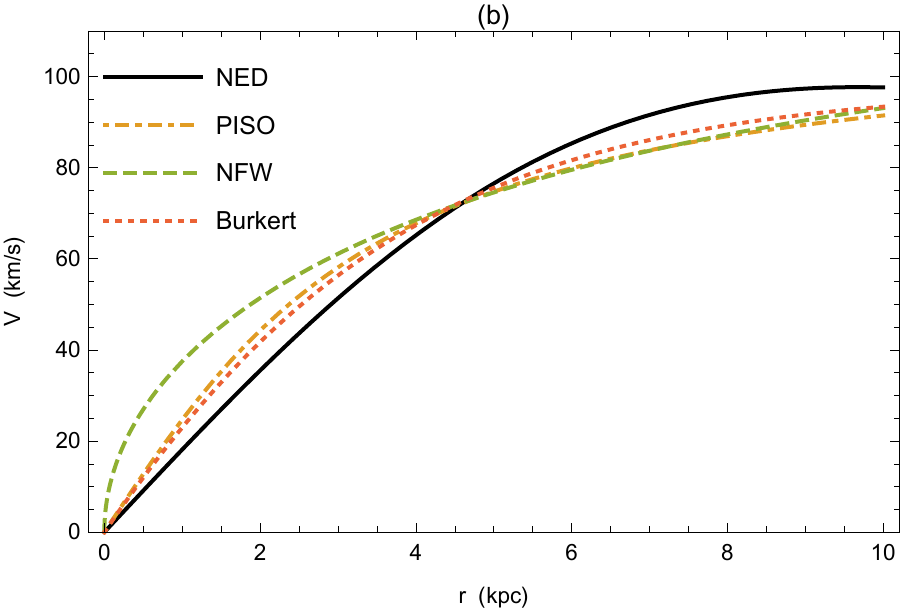}
\end{tabular}
\caption{This figure shows a comparison of rotation curves for NED, PISO, NFW and Burkert models i.e. Eqs. \eqref{velrotnon}, \eqref{velpiso}, \eqref{velnfw} and \eqref{velbur}. In (a) we plot all models with the same parameters values ($\mu_{DM}=92.47$ $M_\odot/$pc$^2$ and $r_i=2.5$ kpc, $i$ for each one of the models). In (b) we have chosen parameters values of NED, NFW and Burkert in order to the rotation curves almost resemble the one of PISO.} 
\label{fig2}
\end{figure}

Let as first  make a comparison of rotation curves associated with PISO, NFW and Burkert (See Appendix \ref{Ap}) versus NED. 
Here we do not considere WaveDM model. 
The comparison between these three models and the NED rotation curves is shown in Fig. \ref{fig2}. 
In Fig. \ref{fig2}(a) we have plotted all the DM models with the same parameters values just to see their behavior. Given that $\mu_{DM}$ is the same for all the four models their different behavior came alone from the dimensionless function $\hat{V}(r/r_i)$ times $r_i$, even do $r_i$ values are the same. 
In Fig. \ref{fig2}(b) we tried to reproduce almost the same behavior of the PISO rotation curve which has been produced with $\mu_{DM}=92.47$ $M_\odot/$pc$^2$ and $r_p=2.5$ kpc, by adjusting the parameters values of the other three models: NED: $\mu_{DM}=59.18$ $M_\odot/$pc$^2$ and $\sqrt\theta=3.2$ kpc; NFW: $\mu_{DM}=55.48$ $M_\odot/$pc$^2$ and $r_{n}=20$ kpc; and Burkert: $\mu_{DM}=166.45$ $M_\odot/$pc$^2$ and $r_b=4.8$ kpc. It has been found that $\mu_{DM}$ is almost constant for cored mass  models\cite{2009MNRAS.397.1169D}.

%%%%%%%%%%%%%%%%%%%%%%%%%%%%%%%%%%%%%%%%
\subsection{Constraints with LSB galaxies}
%%%%%%%%%%%%%%%%%%%%%%%%%%%%%%%%%%%%%%%%%

The main goal of this section, is to compare the results of galaxy rotation curves, comparing the NED parameters fit with parameters fit given by the four most successful models studied in literature which are PISO, NFW, Burkert and WaveDM densities profiles (See Appendix \ref{Ap} for details of the formulae used to fit observations).

In this sense, it is necessary to obtain the best fit, by maximizing the likelihood $\mathcal{L}(\Theta)\varpropto\exp[-\chi^2(\Theta)/2]$, where
\begin{equation}
\chi^{2}(\Theta)=\sum_{j=1}^{N}\left(\frac{V_{\rm obs}^j-V_{{\rm DM} }(r_j,\Theta)}{\delta V_{{\rm obs}}^j}\right)^{2},
\label{chi2Eq}
\end{equation}
here $V^i_{\rm obs}$ and $\delta V^i_{\rm obs}$ is the observed velocity and its corresponding uncertainty at the 
observed 
radial distance $r_j$.
$\Theta$ corresponds to the free parameters of DM model. 
The reduced $\chi^2$ is defined by $\chi^2_{\rm red}=\chi^2/(N-p)$ where $N$ is the total number of data and $p$ is the number of free parameters \cite{numrecip}. Errors in the estimated parameters were computed using the covariance matrix as is described in Ref. \cite{numrecip}.

For the DM models analyzed in this work we have two parameters: a scale length, $r_{i}$, and the density at the center of the galaxy, $\rho_i$; $i$ is for PISO, NFW, Burkert, WaveDM or NED models, 
Eqs. \eqref{PIP}, \eqref{NFW}, \eqref{Burk}, \eqref{rhoNED} and \eqref{eq:WaveDM}. 
As we have already say, we defined in Eq. \eqref{muDM}, the surface density, $\mu_{DM}$, which has turned out to be a very important quantity, characterizing the dark matter models in galaxies \cite{2009MNRAS.397.1169D}. 

%%%%%%%%%%%%%%%%%%%%%%%%
% Sample of LSB galaxies
\begin{center}
\begin{table*}
\ra{1.3}
\begin{center}
\begin{tabular}{@{}l l r r r r r r r r r r r @ {}}\toprule
    \hline \hline
    \multicolumn{12}{c}{Galaxy sample} \\
    \hline 
    \multicolumn{12}{c}{Observational data}\\
   \hline 
\multicolumn{1}{c}{Galaxy} &
\multicolumn{1}{c}{Morphology }&
\multicolumn{1}{c}{$D$ } &
\multicolumn{1}{c}{$V_{\text{hel}}$} &
\multicolumn{1}{c}{$M_{\text{abs}}(B)$}& 
\multicolumn{1}{c}{$R_{\text{max}}$}& 
\multicolumn{1}{c}{$V_{\text{max}}$}\\
\multicolumn{1}{c}{ }& 
\multicolumn{1}{c}{ }& 
\multicolumn{1}{c}{(Mpc)}& 
\multicolumn{1}{c}{(km/s) }& 
\multicolumn{1}{c}{(mag) }& 
\multicolumn{1}{c}{(kpc) }& 
\multicolumn{1}{c}{(km/s) }\\
\multicolumn{1}{c}{(1)}& 
\multicolumn{1}{c}{(2)}& 
\multicolumn{1}{c}{(3)}& 
\multicolumn{1}{c}{(4)}& 
\multicolumn{1}{c}{(5)}& 
\multicolumn{1}{c}{(6)}& 
\multicolumn{1}{c}{(7)}\\
 
\cmidrule[0.4pt](r{0.125em}){1-1}
\cmidrule[0.4pt](lr{0.125em}){2-2}
\cmidrule[0.4pt](lr{0.125em}){3-3}
\cmidrule[0.4pt](lr{0.125em}){4-4}
\cmidrule[0.4pt](lr{0.125em}){5-5}
\cmidrule[0.4pt](lr{0.125em}){6-6}
\cmidrule[0.4pt](lr{0.125em}){7-7}
\cmidrule[0.4pt](lr{0.125em}){8-8}
\cmidrule[0.4pt](lr{0.125em}){9-9}
\cmidrule[0.4pt](lr{0.125em}){10-10}
\cmidrule[0.4pt](lr{0.125em}){11-11}
\cmidrule[0.4pt](lr{0.125em}){12-12}
(1) ESO 0140040 & Spiral & 212 & 16064 & -21.6 & 29.2 & 263 \\ 
(2) ESO 0840411 & Edge-on & 80 & 6200 & -18.1 & 8.9 & 61 \\ 
(3) ESO 1200211 & Fuzzy Magellanic bar & 15 & 1314 & -15.6 & 3.5 & 25 \\ 
(4) ESO 1870510 & Irregular spiral, floculent & 18 & 1410 & -16.5 & 3.0 & 40 \\ 
(5) ESO 2060140 & Spiral & 60 & 4704 & -19.2 & 11.6 & 118 \\ 
(6) ESO 3020120 & Spiral, hint of bar? & 69 & 5311 & -19.1 & 11.0 & 86 \\ 
(7) ESO 3050090 & Barred spiral & 11 & 1019 & -17.3 & 4.8 & 54 \\ 
(8) ESO 4250180 & Spiral & 86 & 6637 & -20.5 & 14.4 & 145 \\ 
(9) ESO 4880049 & Inclined Magellanic bar & 22 & 1800 & -16.8 & 6.0 & 97 \\ 
(10) F730-V1 & Spiral & 144 & 10714 & -19.2 & 11.9 & 145 \\ 
(11) UGC 4115 & Fuzzy & 3.2 & 343 & -12.4 & 1.0 & 40 \\ 
(12) UGC 11454 & Fuzzy spiral, small core & 91 & 6628 & -18.6 & 11.9 & 152 \\ 
(13) UGC 11557 & Fuzzy spiral, small core & 22 & 1390 & -20.0 & 6.2 & 95 \\ 
(14) UGC 11583 & Faint Magellanic bar & 5 & 128 & -14.0 & 1.5 & 36 \\ 
(15) UGC 11616 & Fuzzy, irregular & 73 & 5244 & -20.3 & 9.6 & 143 \\ 
(16) UGC 11648 & Irregular & 48 & 3350 & -21.0 & 12.7 & 145 \\ 
(17) UGC 11748 & Irregular, bright core/bar? & 73 & 5265 & -22.9 & 21.0 & 242 \\ 
(18) UGC 11819 & Fuzzy & 60 & 4261 & -20.3 & 11.7 & 153 \\ 

 \bottomrule
\hline \hline
\end{tabular}
\end{center}
\caption{Sample of LSB galaxies without photometry as given in Ref. \cite{deBlok/etal:2001}. In column (1) we give the name of each galaxy;  in column (2) we give the morphology of each galaxy;
column (3) is the distance to the galaxy in Mpc;  column (4) is the heliocentric systemic velocity;  column (5) is the absolute magnitude in B-band of each galaxy;  column (6) is distance in kpc where the maximum velocity is reach;  column (7) is the maximum velocity of each rotation curve.}\label{tab:sample}
\end{table*}
\end{center}
%%%%%%%%%%%%%%%%%%%%%%%%
%

%%%%%%%%%%%%%%%%%%%%%%%%
% Fitting results PISO
\begin{center}
\begin{table*}
\ra{1.3}
\begin{center}
\begin{tabular}{@{}l r r r r r r r r r r @ {}}\toprule
    \hline \hline
    \multicolumn{7}{c}{PISO DM Model}\\
    \hline 
   \multicolumn{7}{c}{Best Fitting Parameters}\\
   \hline 
\multicolumn{1}{c}{Galaxy} &
\multicolumn{1}{c}{$\rho_{p}$ }&
\multicolumn{1}{c}{$r_{p}$ } &
\multicolumn{1}{c}{$\mu_{DM}$} &
\multicolumn{1}{c}{$M_{DM}(300$ pc$)$}& 
\multicolumn{1}{c}{$M_{DM}(r_{\text{max}})$}& 
\multicolumn{1}{c}{$\chi^2_{\text{red}}$}\\
\multicolumn{1}{c}{ }& 
\multicolumn{1}{c}{($10^{-3}M_\odot/$pc$^3$)}& 
\multicolumn{1}{c}{(kpc) }& 
\multicolumn{1}{c}{($M_\odot/$pc$^2$) }& 
\multicolumn{1}{c}{($10^{7}M_\odot$)}& 
\multicolumn{1}{c}{($10^{11}M_\odot$)}& 
\multicolumn{1}{c}{ }\\
\multicolumn{1}{c}{(1)}& 
\multicolumn{1}{c}{(2)}& 
\multicolumn{1}{c}{(3)}& 
\multicolumn{1}{c}{(4)}& 
\multicolumn{1}{c}{(5)}& 
\multicolumn{1}{c}{(6)}& 
\multicolumn{1}{c}{(7)}\\
 
\cmidrule[0.4pt](r{0.25em}){1-1}
\cmidrule[0.4pt](lr{0.25em}){2-2}
\cmidrule[0.4pt](lr{0.25em}){3-3}
\cmidrule[0.4pt](lr{0.25em}){4-4}
\cmidrule[0.4pt](lr{0.25em}){5-5}
\cmidrule[0.4pt](lr{0.25em}){6-6}
\cmidrule[0.4pt](lr{0.25em}){7-7}
(1) ESO 0140040 & 249.301$\pm$100.439 & 2.559$\pm$0.657 & 11.798 & 2.797 & 5.374 & 0.180 \\ 
(2) ESO 0840411 & 5.235$\pm$2.031 & 6.376$\pm$3.308 & 0.617 & 0.059 & 0.087 & 0.050 \\ 
(3) ESO 1200211 & 45.952$\pm$45.122 & 0.573$\pm$0.382 & 0.487 & 0.448 & 0.005 & 0.078 \\ 
(4) ESO 1870510 & 54.231$\pm$29.375 & 0.967$\pm$0.448 & 0.970 & 0.580 & 0.012 & 0.028 \\ 
(5) ESO 2060140 & 233.076$\pm$79.144 & 1.164$\pm$0.239 & 5.016 & 2.536 & 0.408 & 0.106 \\ 
(6) ESO 3020120 & 54.204$\pm$26.987 & 1.895$\pm$0.707 & 1.900 & 0.604 & 0.216 & 0.034 \\ 
(7) ESO 3050090 & 27.322$\pm$13.025 & 2.091$\pm$1.005 & 1.057 & 0.305 & 0.038 & 0.048 \\ 
(8) ESO 4250180 & 30.192$\pm$37.443 & 4.398$\pm$4.190 & 2.456 & 0.341 & 0.686 & 0.095 \\ 
(9) ESO 4880049 & 102.348$\pm$38.523 & 1.622$\pm$0.469 & 3.071 & 1.134 & 0.138 & 0.017 \\ 
(10) F730-V1 & 214.024$\pm$69.305 & 1.472$\pm$0.304 & 5.828 & 2.362 & 0.587 & 0.083 \\ 
(11) UGC 4115 & 149.748$\pm$63.661 & 0.931$\pm$0.765 & 2.579 & 1.595 & 0.004 & 0.004 \\ 
(12) UGC 11454 & 151.154$\pm$29.156 & 1.926$\pm$0.246 & 5.385 & 1.685 & 0.668 & 0.382 \\ 
(13) UGC 11557 & 15.453$\pm$6.272 & 5.378$\pm$3.683 & 1.537 & 0.174 & 0.093 & 0.051 \\ 
(14) UGC 11583 & 119.175$\pm$72.692 & 0.633$\pm$0.355 & 1.394 & 1.191 & 0.005 & 0.106 \\ 
(15) UGC 11616 & 199.314$\pm$43.504 & 1.498$\pm$0.223 & 5.521 & 2.201 & 0.431 & 0.171 \\ 
(16) UGC 11648 & 105.879$\pm$19.255 & 1.948$\pm$0.235 & 3.814 & 1.181 & 0.511 & 3.739 \\ 
(17) UGC 11748 & 8165.259$\pm$3307.033 & 0.367$\pm$0.078 & 55.482 & 67.067 & 2.857 & 5.719 \\ 
(18) UGC 11819 & 88.880$\pm$14.071 & 2.933$\pm$0.387 & 4.821 & 0.999 & 0.769 & 0.316 \\ 

 \bottomrule
\hline \hline
\end{tabular}
\end{center}
\caption{Best fitting parameters for PISO DM model and derived quantities.}\label{tab:PISO}
\end{table*}
\end{center}
%%%%%%%%%%%%%%%%%%%%%%%%
%

%%%%%%%%%%%%%%%%%%%%%%%%
% Fitting results NFW
\begin{center}
\begin{table*}
\ra{1.3}
\begin{center}
\begin{tabular}{@{}l r r r r r r r r r r @ {}}\toprule
    \hline \hline
    \multicolumn{7}{c}{NFW DM Model}\\
    \hline 
   \multicolumn{7}{c}{Best Fitting Parameters}\\
   \hline 
\multicolumn{1}{c}{Galaxy} &
\multicolumn{1}{c}{$\rho_{n}$ }&
\multicolumn{1}{c}{$r_{n}$ } &
\multicolumn{1}{c}{$\mu_{DM}$} &
\multicolumn{1}{c}{$M_{DM}(300$ pc$)$}& 
\multicolumn{1}{c}{$M_{DM}(r_{\text{max}})$}& 
\multicolumn{1}{c}{$\chi^2_{\text{red}}$}\\
\multicolumn{1}{c}{ }& 
\multicolumn{1}{c}{($10^{-3}M_\odot/$pc$^3$)}& 
\multicolumn{1}{c}{(kpc) }& 
\multicolumn{1}{c}{($M_\odot/$pc$^2$) }& 
\multicolumn{1}{c}{($10^{7}M_\odot$)}& 
\multicolumn{1}{c}{($10^{11}M_\odot$)}& 
\multicolumn{1}{c}{ }\\
\multicolumn{1}{c}{(1)}& 
\multicolumn{1}{c}{(2)}& 
\multicolumn{1}{c}{(3)}& 
\multicolumn{1}{c}{(4)}& 
\multicolumn{1}{c}{(5)}& 
\multicolumn{1}{c}{(6)}& 
\multicolumn{1}{c}{(7)}\\
 
\cmidrule[0.4pt](r{0.25em}){1-1}
\cmidrule[0.4pt](lr{0.25em}){2-2}
\cmidrule[0.4pt](lr{0.25em}){3-3}
\cmidrule[0.4pt](lr{0.25em}){4-4}
\cmidrule[0.4pt](lr{0.25em}){5-5}
\cmidrule[0.4pt](lr{0.25em}){6-6}
\cmidrule[0.4pt](lr{0.25em}){7-7}
(1) ESO 0140040 & 25.478$\pm$11.506 & 16.148$\pm$4.602 & 7.609 & 22.701 & 5.392 & 0.142 \\ 
(2) ESO 0840411 & 0.011$\pm$0.023 & 1157.493$\pm$2464.316 & 0.233 & 0.713 & 0.071 & 1.728 \\ 
(3) ESO 1200211 & 2.419$\pm$5.244 & 5.705$\pm$9.200 & 0.255 & 0.729 & 0.006 & 0.239 \\ 
(4) ESO 1870510 & 0.753$\pm$3.829 & 31.828$\pm$150.341 & 0.443 & 1.338 & 0.013 & 0.057 \\ 
(5) ESO 2060140 & 19.886$\pm$9.262 & 8.110$\pm$2.327 & 2.983 & 8.688 & 0.415 & 0.420 \\ 
(6) ESO 3020120 & 2.622$\pm$3.451 & 19.720$\pm$20.010 & 0.956 & 2.865 & 0.230 & 0.328 \\ 
(7) ESO 3050090 & 3.153$\pm$2.380 & 10.000$\pm$5.829 & 0.583 & 1.714 & 0.029 & 0.654 \\ 
(8) ESO 4250180 & 0.525$\pm$3.747 & 119.446$\pm$770.596 & 1.159 & 3.531 & 0.756 & 0.013 \\ 
(9) ESO 4880049 & 10.366$\pm$6.610 & 10.000$\pm$4.634 & 1.917 & 5.635 & 0.130 & 0.608 \\ 
(10) F730-V1 & 18.747$\pm$7.817 & 10.000$\pm$2.709 & 3.467 & 10.191 & 0.584 & 1.455 \\ 
(11) UGC 4115 & 5.093$\pm$6.316 & 10.00$\pm$11.428 & 0.942 & 2.769 & 0.003 & 0.939 \\ 
(12) UGC 11454 & 20.351$\pm$4.611 & 10.000$\pm$1.447 & 3.764 & 11.063 & 0.634 & 4.780 \\ 
(13) UGC 11557 & 4.738$\pm$2.736 & 10.000$\pm$4.128 & 0.876 & 2.576 & 0.061 & 3.265 \\ 
(14) UGC 11583 & 3.629$\pm$5.640 & 10.000$\pm$13.774 & 0.671 & 1.973 & 0.005 & 0.833 \\ 
(15) UGC 11616 & 18.362$\pm$6.408 & 10.000$\pm$2.393 & 3.396 & 9.982 & 0.434 & 1.683 \\ 
(16) UGC 11648 & 14.585$\pm$2.799 & 10.000$\pm$1.216 & 2.697 & 7.929 & 0.486 & 3.031 \\ 
(17) UGC 11748 & 498.342$\pm$92.553 & 3.191$\pm$0.288 & 29.411 & 79.738 & 2.370 & 3.546 \\ 
(18) UGC 11819 & 20.272$\pm$5.070 & 10.000$\pm$1.674 & 3.749 & 11.020 & 0.613 & 4.473 \\

\bottomrule
\hline \hline
\end{tabular}
\end{center}
\caption{Best fitting parameters for NFW dark matter model and derived quantities.}\label{tab:NFW}
\end{table*}
\end{center}
%%%%%%%%%%%%%%%%%%%%%%%%
%

%%%%%%%%%%%%%%%%%%%%%%%%
% Fitting results Burkert
\begin{center}
\begin{table*}
\ra{1.3}
\begin{center}
\begin{tabular}{@{}l r r r r r r r r r r @ {}}\toprule
    \hline \hline
    \multicolumn{7}{c}{Burkert DM Model}\\
    \hline 
   \multicolumn{7}{c}{Best Fitting Parameters}\\
   \hline 
\multicolumn{1}{c}{Galaxy} &
\multicolumn{1}{c}{$\rho_{b}$ }&
\multicolumn{1}{c}{$r_{b}$ } &
\multicolumn{1}{c}{$\mu_{DM}$} &
\multicolumn{1}{c}{$M_{DM}(300$ pc$)$}& 
\multicolumn{1}{c}{$M_{DM}(r_{\text{max}})$}& 
\multicolumn{1}{c}{$\chi^2_{\text{red}}$}\\
\multicolumn{1}{c}{ }& 
\multicolumn{1}{c}{($10^{-3}M_\odot/$pc$^3$)}& 
\multicolumn{1}{c}{(kpc) }& 
\multicolumn{1}{c}{($M_\odot/$pc$^2$) }& 
\multicolumn{1}{c}{($10^{7}M_\odot$)}& 
\multicolumn{1}{c}{($10^{11}M_\odot$)}& 
\multicolumn{1}{c}{ }\\
\multicolumn{1}{c}{(1)}& 
\multicolumn{1}{c}{(2)}& 
\multicolumn{1}{c}{(3)}& 
\multicolumn{1}{c}{(4)}& 
\multicolumn{1}{c}{(5)}& 
\multicolumn{1}{c}{(6)}& 
\multicolumn{1}{c}{(7)}\\
 
\cmidrule[0.4pt](r{0.25em}){1-1}
\cmidrule[0.4pt](lr{0.25em}){2-2}
\cmidrule[0.4pt](lr{0.25em}){3-3}
\cmidrule[0.4pt](lr{0.25em}){4-4}
\cmidrule[0.4pt](lr{0.25em}){5-5}
\cmidrule[0.4pt](lr{0.25em}){6-6}
\cmidrule[0.4pt](lr{0.25em}){7-7}
(1) ESO 0140040 & 174.894$\pm$37.361 & 5.973$\pm$0.807 & 19.320 & 1.903 & 4.811 & 0.556 \\ 
(2) ESO 0840411 & 5.848$\pm$2.402 & 10.423$\pm$6.001 & 1.127 & 0.065 & 0.088 & 0.034 \\ 
(3) ESO 1200211 & 47.651$\pm$39.331 & 1.010$\pm$0.529 & 0.890 & 0.420 & 0.005 & 0.070 \\ 
(4) ESO 1870510 & 59.656$\pm$28.150 & 1.603$\pm$0.628 & 1.768 & 0.580 & 0.012 & 0.031 \\ 
(5) ESO 2060140 & 190.003$\pm$46.654 & 2.480$\pm$0.336 & 8.713 & 1.954 & 0.363 & 0.193 \\ 
(6) ESO 3020120 & 56.685$\pm$23.031 & 3.372$\pm$0.960 & 3.535 & 0.598 & 0.200 & 0.006 \\ 
(7) ESO 3050090 & 30.689$\pm$14.441 & 3.376$\pm$1.581 & 1.916 & 0.324 & 0.038 & 0.042 \\ 
(8) ESO 4250180 & 30.713$\pm$30.272 & 7.695$\pm$5.812 & 4.371 & 0.337 & 0.676 & 0.111 \\ 
(9) ESO 4880049 & 110.947$\pm$37.263 & 2.729$\pm$0.669 & 5.599 & 1.151 & 0.133 & 0.046 \\ 
(10) F730-V1 & 191.278$\pm$50.734 & 2.931$\pm$0.460 & 10.370 & 1.997 & 0.533 & 0.569 \\ 
(11) UGC 4115 & 167.708$\pm$81.223 & 1.543$\pm$1.520 & 4.787 & 1.621 & 0.004 & 0.002 \\ 
(12) UGC 11454 & 138.988$\pm$20.518 & 3.705$\pm$0.342 & 9.524 & 1.476 & 0.629 & 0.981 \\ 
(13) UGC 11557 & 17.308$\pm$7.708 & 8.880$\pm$7.056 & 2.843 & 0.191 & 0.094 & 0.046 \\ 
(14) UGC 11583 & 138.330$\pm$81.382 & 0.990$\pm$0.517 & 2.533 & 1.213 & 0.005 & 0.111 \\ 
(15) UGC 11616 & 195.956$\pm$33.714 & 2.751$\pm$0.304 & 9.971 & 2.035 & 0.392 & 0.326 \\ 
(16) UGC 11648 & 81.798$\pm$10.504 & 4.217$\pm$0.352 & 6.380 & 0.876 & 0.507 & 5.450 \\ 
(17) UGC 11748 & 1438.345$\pm$179.750 & 1.932$\pm$0.115 & 51.384 & 14.376 & 2.216 & 2.164 \\ 
(18) UGC 11819 & 98.280$\pm$13.412 & 4.843$\pm$0.531 & 8.803 & 1.060 & 0.724 & 0.189 \\ 

 \bottomrule
\hline \hline
\end{tabular}
\end{center}
\caption{Best fitting parameters for Burkert DM model and derived quantities.}\label{tab:Burkert}
\end{table*}
\end{center}
%%%%%%%%%%%%%%%%%%%%%%%%
%

%%%%%%%%%%%%%%%%%%%%%%%%
% Fitting results NED
\begin{center}
\begin{table*}
\ra{1.3}
\begin{center}
\begin{tabular}{@{}l r r r r r r r r r r @ {}}\toprule
    \hline \hline
   \multicolumn{7}{c}{NED DM Model}\\
    \hline 
   \multicolumn{7}{c}{Best Fitting Parameters}\\
   \hline 
\multicolumn{1}{c}{Galaxy} &
\multicolumn{1}{c}{$\rho_{\text{NED}}$ }&
\multicolumn{1}{c}{$\sqrt\theta$ } &
\multicolumn{1}{c}{$\mu_{DM}$} &
\multicolumn{1}{c}{$M_{DM}(300$ pc$)$}& 
\multicolumn{1}{c}{$M_{DM}(r_{\text{max}})$}& 
\multicolumn{1}{c}{$\chi^2_{\text{red}}$}\\
\multicolumn{1}{c}{ }& 
\multicolumn{1}{c}{($10^{-3}M_\odot/$pc$^3$)}& 
\multicolumn{1}{c}{(kpc) }& 
\multicolumn{1}{c}{($M_\odot/$pc$^2$) }& 
\multicolumn{1}{c}{($10^{7}M_\odot$)}& 
\multicolumn{1}{c}{($10^{11}M_\odot$)}& 
\multicolumn{1}{c}{ }\\
\multicolumn{1}{c}{(1)}& 
\multicolumn{1}{c}{(2)}& 
\multicolumn{1}{c}{(3)}& 
\multicolumn{1}{c}{(4)}& 
\multicolumn{1}{c}{(5)}& 
\multicolumn{1}{c}{(6)}& 
\multicolumn{1}{c}{(7)}\\
 
\cmidrule[0.4pt](r{0.25em}){1-1}
\cmidrule[0.4pt](lr{0.25em}){2-2}
\cmidrule[0.4pt](lr{0.25em}){3-3}
\cmidrule[0.4pt](lr{0.25em}){4-4}
\cmidrule[0.4pt](lr{0.25em}){5-5}
\cmidrule[0.4pt](lr{0.25em}){6-6}
\cmidrule[0.4pt](lr{0.25em}){7-7}
(1) ESO 0140040 & 74.830$\pm$7.787 & 4.795$\pm$0.278 & 6.636 & 0.846 & 3.674 & 2.678 \\ 
(2) ESO 0840411 & 4.858$\pm$1.557 & 4.141$\pm$1.516 & 0.372 & 0.055 & 0.084 & 0.083 \\ 
(3) ESO 1200211 & 23.382$\pm$13.305 & 0.693$\pm$0.221 & 0.300 & 0.257 & 0.003 & 0.160 \\ 
(4) ESO 1870510 & 37.574$\pm$11.172 & 0.902$\pm$0.192 & 0.627 & 0.418 & 0.011 & 0.115 \\ 
(5) ESO 2060140 & 76.224$\pm$12.544 & 2.004$\pm$0.151 & 2.826 & 0.859 & 0.273 & 2.311 \\ 
(6) ESO 3020120 & 33.054$\pm$9.507 & 2.202$\pm$0.364 & 1.346 & 0.373 & 0.157 & 0.172 \\ 
(7) ESO 3050090 & 23.122$\pm$8.218 & 1.592$\pm$0.448 & 0.681 & 0.260 & 0.034 & 0.082 \\ 
(8) ESO 4250180 & 16.372$\pm$8.696 & 4.752$\pm$1.960 & 1.439 & 0.185 & 0.647 & 0.244 \\ 
(9) ESO 4880049 & 66.721$\pm$15.229 & 1.599$\pm$0.220 & 1.974 & 0.751 & 0.115 & 0.408 \\ 
(10) F730-V1 & 80.499$\pm$17.124 & 2.271$\pm$0.253 & 3.381 & 0.908 & 0.419 & 5.403 \\ 
(11) UGC 4115 & 144.243$\pm$50.285 & 0.556$\pm$0.340 & 1.484 & 1.562 & 0.004 & 0.006 \\ 
(12) UGC 11454 & 61.686$\pm$5.589 & 2.724$\pm$0.137 & 3.107 & 0.696 & 0.545 & 6.318 \\ 
(13) UGC 11557 & 14.532$\pm$4.656 & 3.360$\pm$1.691 & 0.903 & 0.164 & 0.092 & 0.068 \\ 
(14) UGC 11583 & 99.834$\pm$40.813 & 0.486$\pm$0.143 & 0.898 & 1.067 & 0.004 & 0.054 \\ 
(15) UGC 11616 & 98.889$\pm$11.088 & 1.879$\pm$0.115 & 3.436 & 1.114 & 0.291 & 2.800 \\ 
(16) UGC 11648 & 35.237$\pm$2.594 & 3.179$\pm$0.136 & 2.072 & 0.398 & 0.483 & 11.531 \\ 
(17) UGC 11748 & 327.142$\pm$18.981 & 2.217$\pm$0.056 & 13.412 & 3.690 & 1.587 & 16.596 \\ 
(18) UGC 11819 & 60.543$\pm$5.211 & 2.782$\pm$0.168 & 3.115 & 0.684 & 0.565 & 0.453 \\

\bottomrule
\hline \hline
\end{tabular}
\end{center}
\caption{Best fitting parameters for NED dark matter model and derived quantities.}\label{tab:NED}
\end{table*}
\end{center}
%%%%%%%%%%%%%%%%%%%%%%%%
%

%%%%%%%%%%%%%%%%%%%%%%%%
% Fitting results WaveDM
\begin{center}
\begin{table*}
\ra{1.3}
\begin{center}
\begin{tabular}{@{}l r r r r r r r r r r @ {}}\toprule
    \hline \hline
%    \multicolumn{7}{c}{Galaxy sample} \\
    \multicolumn{7}{c}{WaveDM Model}\\
%    \multicolumn{7}{c}{WaveDM parameters}\\
    \hline 
   \multicolumn{7}{c}{Best Fitting Parameters}\\
   \hline 
\multicolumn{1}{c}{Galaxy} &
\multicolumn{1}{c}{$\rho_{w}$ }&
\multicolumn{1}{c}{$r_{w}$ } &
\multicolumn{1}{c}{$\mu_{DM}$} &
\multicolumn{1}{c}{$M_{DM}(300$ pc$)$}& 
\multicolumn{1}{c}{$M_{DM}(r_{\text{max}})$}& 
\multicolumn{1}{c}{$\chi^2_{\text{red}}$}\\
\multicolumn{1}{c}{ }& 
\multicolumn{1}{c}{($10^{-3}M_\odot/$pc$^3$)}& 
\multicolumn{1}{c}{(kpc) }& 
\multicolumn{1}{c}{($M_\odot/$pc$^2$) }& 
\multicolumn{1}{c}{($10^{7}M_\odot$)}& 
\multicolumn{1}{c}{($10^{11}M_\odot$)}& 
\multicolumn{1}{c}{ }\\
\multicolumn{1}{c}{(1)}& 
\multicolumn{1}{c}{(2)}& 
\multicolumn{1}{c}{(3)}& 
\multicolumn{1}{c}{(4)}& 
\multicolumn{1}{c}{(5)}& 
\multicolumn{1}{c}{(6)}& 
\multicolumn{1}{c}{(7)}\\
 
\cmidrule[0.4pt](r{0.25em}){1-1}
\cmidrule[0.4pt](lr{0.25em}){2-2}
\cmidrule[0.4pt](lr{0.25em}){3-3}
\cmidrule[0.4pt](lr{0.25em}){4-4}
\cmidrule[0.4pt](lr{0.25em}){5-5}
\cmidrule[0.4pt](lr{0.25em}){6-6}
\cmidrule[0.4pt](lr{0.25em}){7-7}
%Insert here each galaxy parameter line
(1) ESO 0140040 & 79.214$\pm$8.781 & 24.751$\pm$1.576 & 36.260 & 0.895 & 3.794 & 2.324 \\ 
(2) ESO 0840411 & 4.903$\pm$1.610 & 22.717$\pm$8.684 & 2.060 & 0.055 & 0.085 & 0.078 \\ 
(3) ESO 1200211 & 25.106$\pm$14.740 & 3.592$\pm$1.190 & 1.668 & 0.275 & 0.004 & 0.132 \\ 
(4) ESO 1870510 & 38.935$\pm$12.248 & 4.793$\pm$1.107 & 3.452 & 0.432 & 0.011 & 0.098 \\ 
(5) ESO 2060140 & 82.166$\pm$13.814 & 10.321$\pm$0.815 & 15.684 & 0.926 & 0.285 & 1.866 \\ 
(6) ESO 3020120 & 34.495$\pm$10.131 & 11.499$\pm$1.992 & 7.336 & 0.389 & 0.163 & 0.116 \\ 
(7) ESO 3050090 & 23.541$\pm$8.625 & 8.603$\pm$2.582 & 3.746 & 0.265 & 0.035 & 0.077 \\ 
(8) ESO 4250180 & 17.341$\pm$10.170 & 24.965$\pm$11.206 & 8.006 & 0.196 & 0.648 & 0.223 \\ 
(9) ESO 4880049 & 69.977$\pm$16.824 & 8.430$\pm$1.254 & 10.910 & 0.787 & 0.117 & 0.334 \\ 
(10) F730-V1 & 87.966$\pm$18.401 & 11.636$\pm$1.277 & 18.930 & 0.992 & 0.433 & 4.428 \\ 
(11) UGC 4115 & 144.859$\pm$51.666 & 3.083$\pm$1.955 & 8.259 & 1.566 & 0.004 & 0.006 \\ 
(12) UGC 11454 & 66.896$\pm$6.449 & 14.024$\pm$0.759 & 17.350 & 0.755 & 0.554 & 5.262 \\ 
(13) UGC 11557 & 25.866$\pm$6.737 & 10.000$\pm$1.747 & 4.784 & 0.291 & 0.065 & 1.078 \\ 
(14) UGC 11583 & 102.235$\pm$43.907 & 2.615$\pm$0.830 & 4.944 & 1.086 & 0.004 & 0.056 \\ 
(15) UGC 11616 & 104.803$\pm$12.177 & 9.795$\pm$0.635 & 18.986 & 1.180 & 0.305 & 2.219 \\ 
(16) UGC 11648 & 37.922$\pm$2.977 & 16.424$\pm$0.767 & 11.519 & 0.428 & 0.486 & 10.683 \\ 
(17) UGC 11748 & 357.566$\pm$21.695 & 11.261$\pm$0.300 & 74.470 & 4.030 & 1.624 & 13.950 \\ 
(18) UGC 11819 & 62.928$\pm$5.724 & 14.740$\pm$0.959 & 17.154 & 0.710 & 0.590 & 0.336 \\ 

\bottomrule
\hline \hline
\end{tabular}
\end{center}
\caption{Best fitting parameters for WaveDM model and derived quantities.}\label{tab:WDM}
\end{table*}
\end{center}
%%%%%%%%%%%%%%%%%%%%%%%%
%

%
We analyze a sample of eighteen high resolution rotation curves of LSB galaxies with no photometry:
an optical rotation curve were available but no optical or H I photometry  \cite{deBlok/etal:2001}.
Accordingly we neglect the visible components, such as gas and stars. 
The sample of analyzed galaxies is given in table  \ref{tab:sample}. See Ref. \cite{deBlok/etal:2001} for technical details.
We remark that in this subsection 
we use units such that $G=1$, velocities are in km/s, and distances are given in kpc.

Results are summarized in Tables \ref{tab:PISO} for PISO model, \ref{tab:NFW} for NFW model, \ref{tab:Burkert} for Burkert model, \ref{tab:NED} for NED model and \ref{tab:WDM} for WaveDM model. 
In those tables we show for each of the studied galaxies the fitting parameters $\rho_i$ and $r_i$ for each one of the five models, columns (2) and (3) and the corresponding 
$\chi_{\text{red}}^2$ is in column (7). Also shown are the estimated fitting errors and the derived quantities like: $\mu_{DM}$, the mass up to 300 pc and the mass up to the outer spatial point measured, $r_{\text{max}}$, columns (4), (5) and (6) respectively. Results for parameters $\rho_i$ and $r_i$ for PISO and NFW models are very similar to those found in Ref.  \cite{deBlok/etal:2001}, except in the estimated errors. This could be due to the method used to fit the data and the method to estimate the fitting errors. As we mentioned before we use the covariance matrix to compute the errors following Ref. \cite{numrecip} were its diagonal elements gives the errors in the estimated parameters. However, some programs do not include the factor $\sqrt{2.30}$ that  Ref. \cite{numrecip} does take into account. In this way the estimated errors in the parameters reported here will diminish a great deal. 

\begin{figure*}[h]
\centering
\begin{tabular}{ccc}
\includegraphics[scale=0.4]{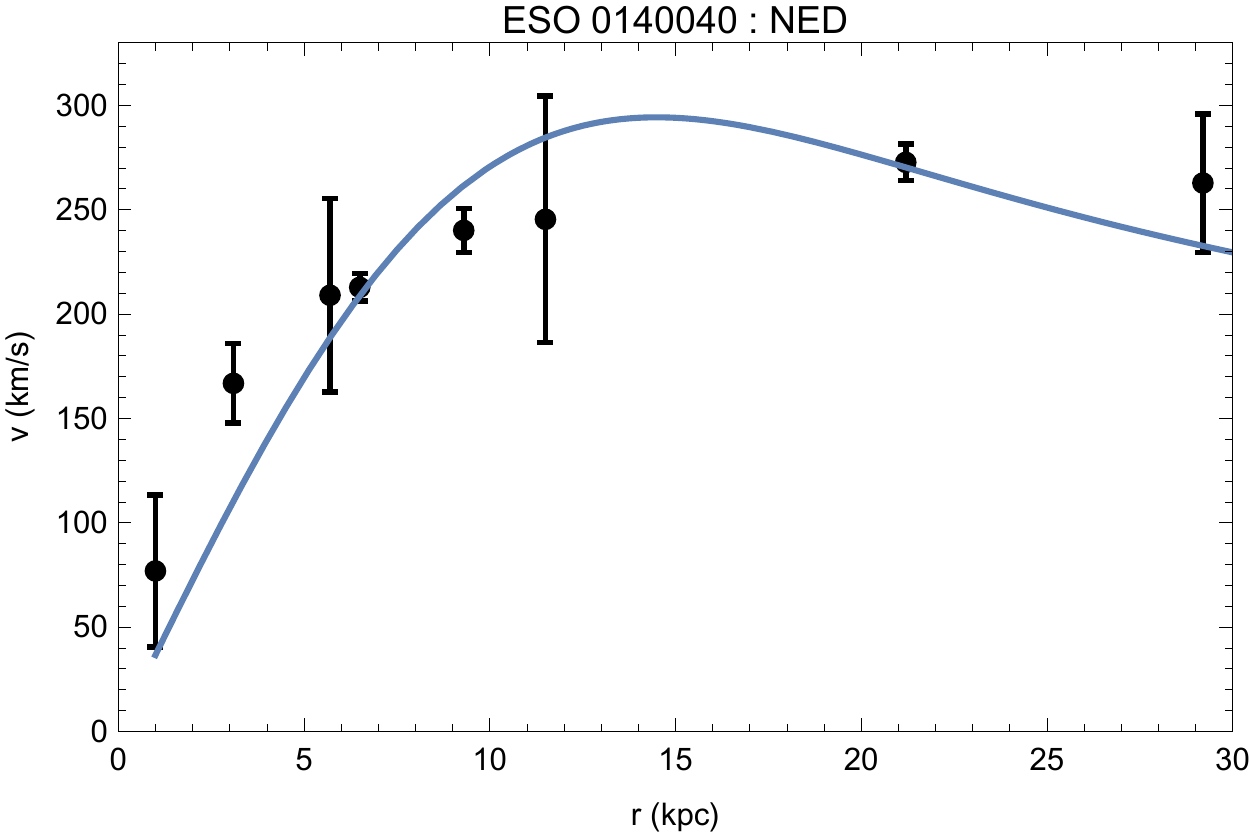} &
\includegraphics[scale=0.4]{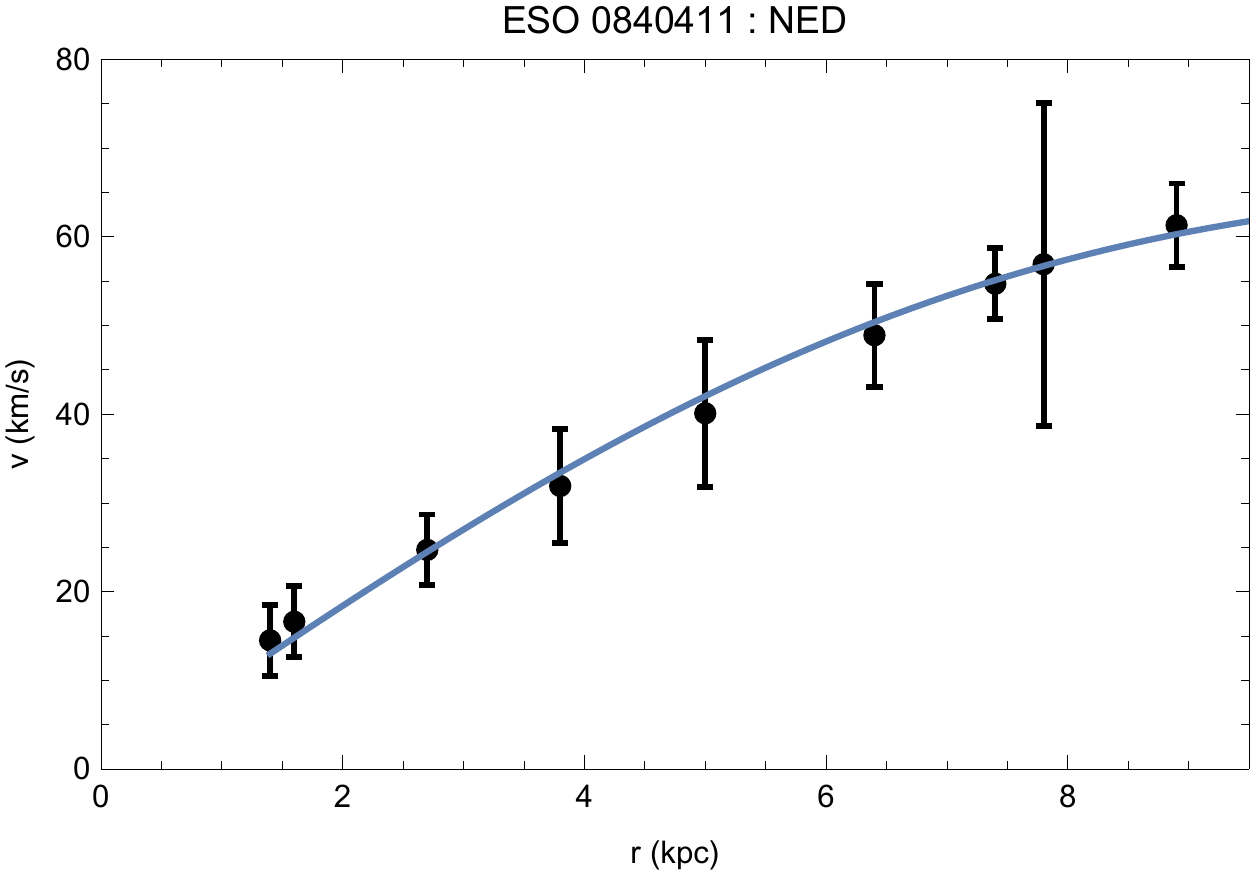} &
\includegraphics[scale=0.4]{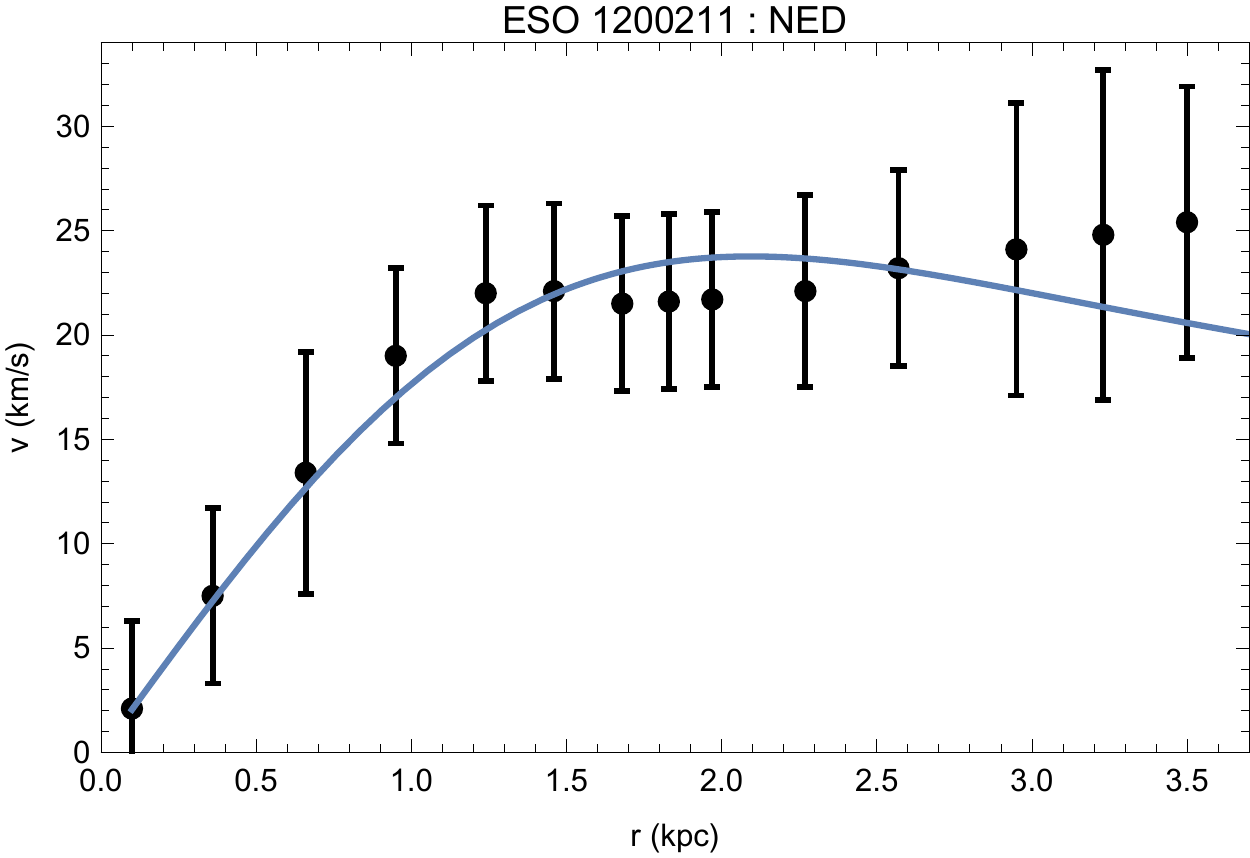} \\  
\includegraphics[scale=0.4]{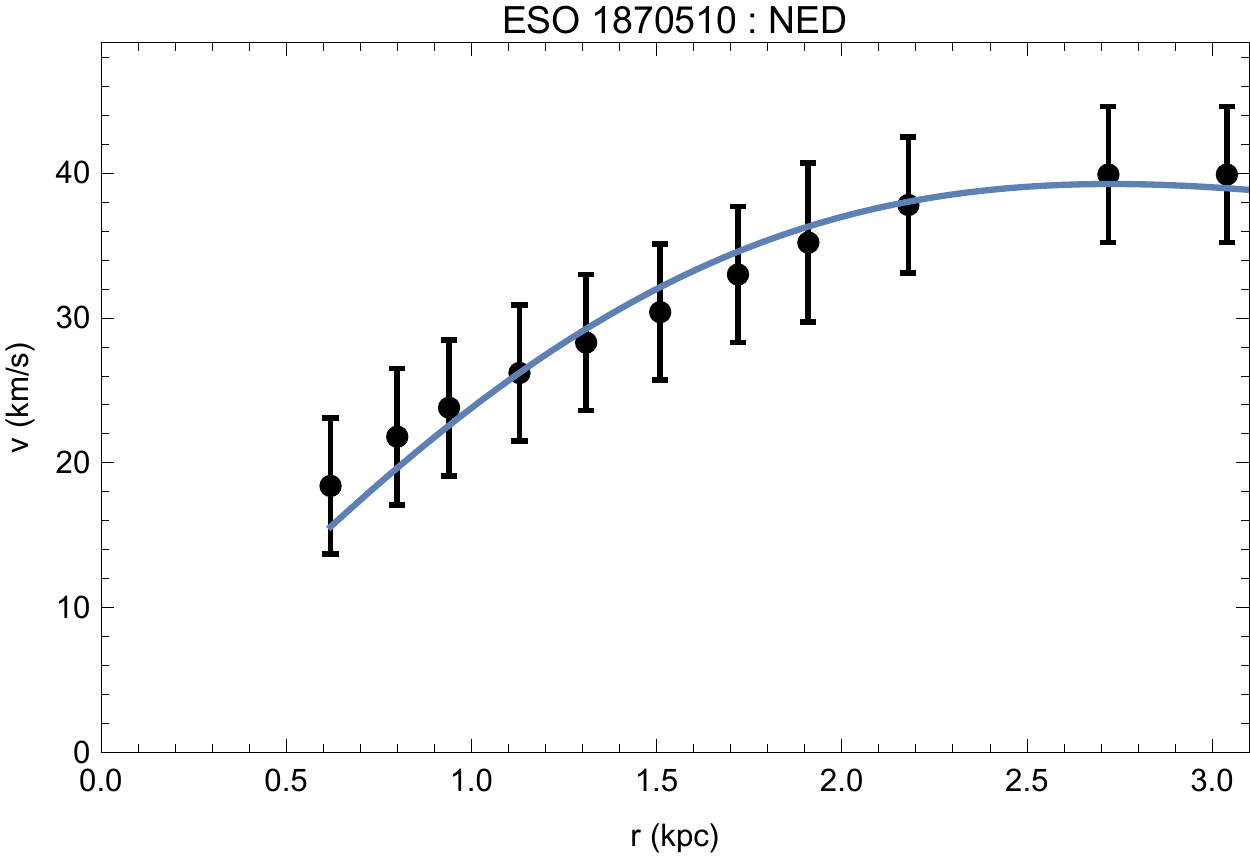} &   
\includegraphics[scale=0.4]{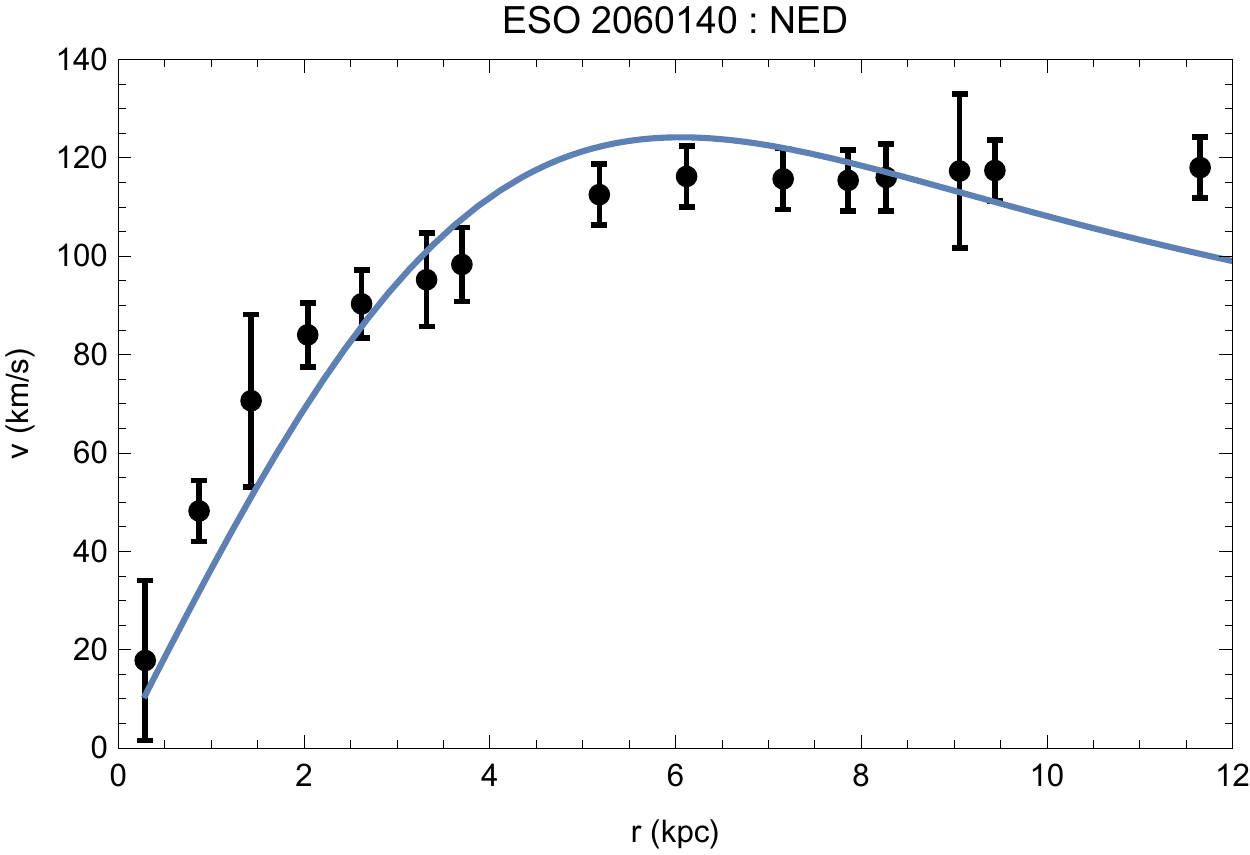} &
\includegraphics[scale=0.4]{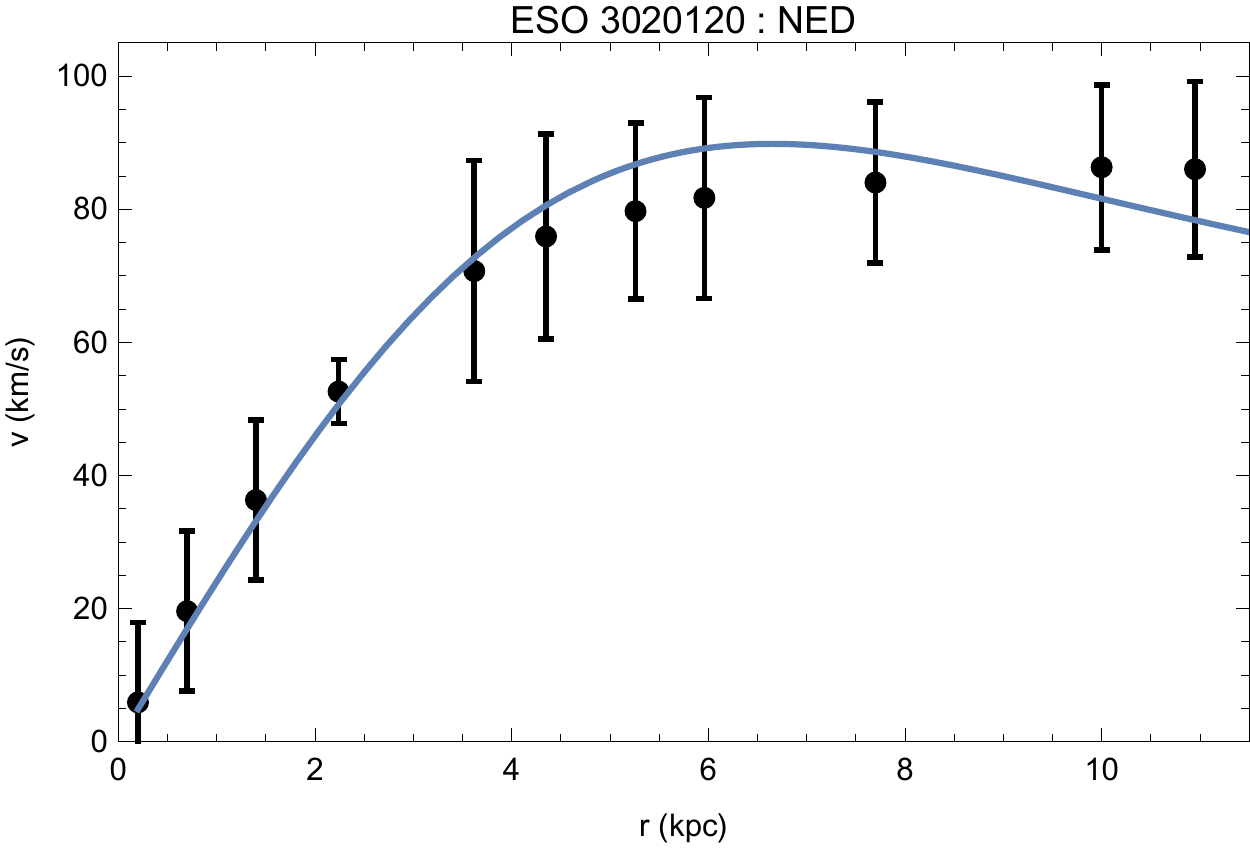} \\   
\includegraphics[scale=0.4]{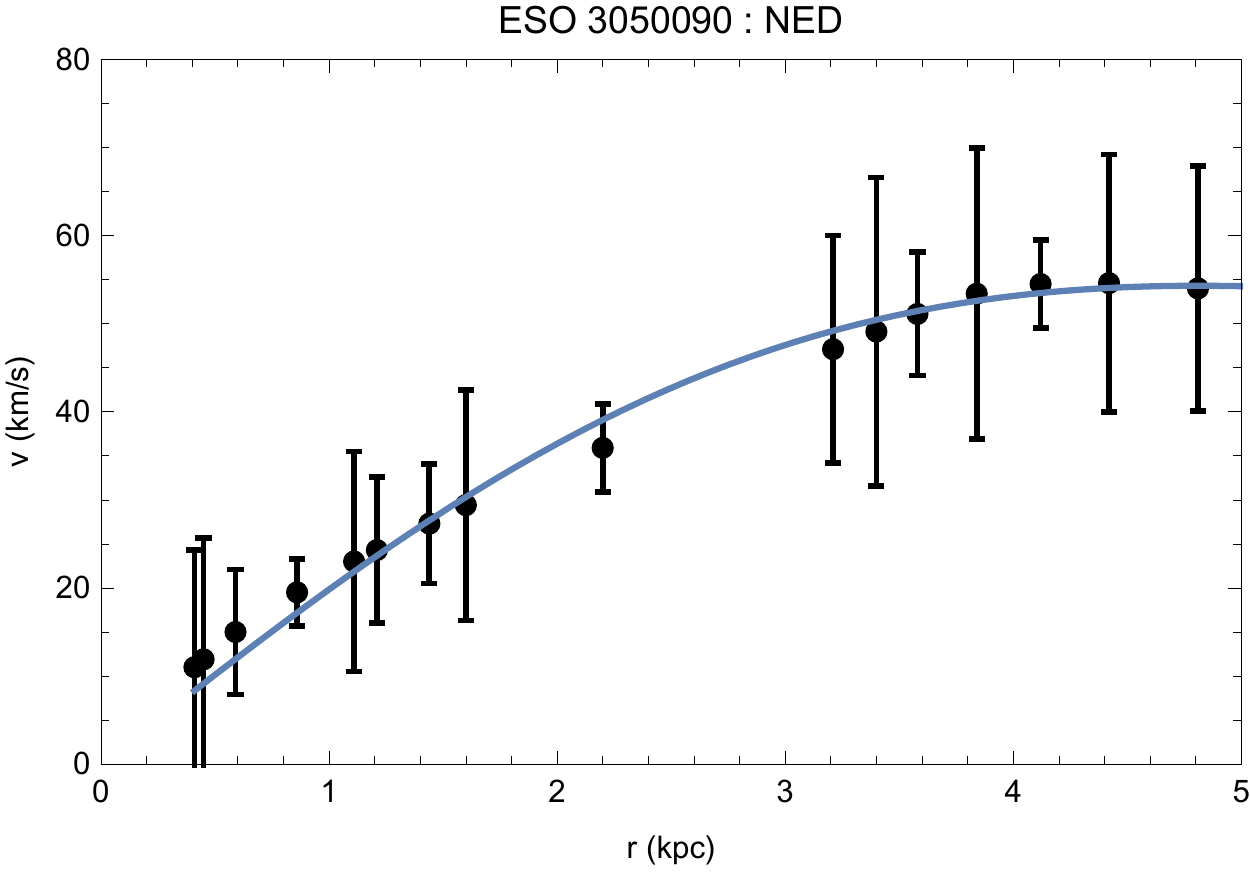} & 
\includegraphics[scale=0.4]{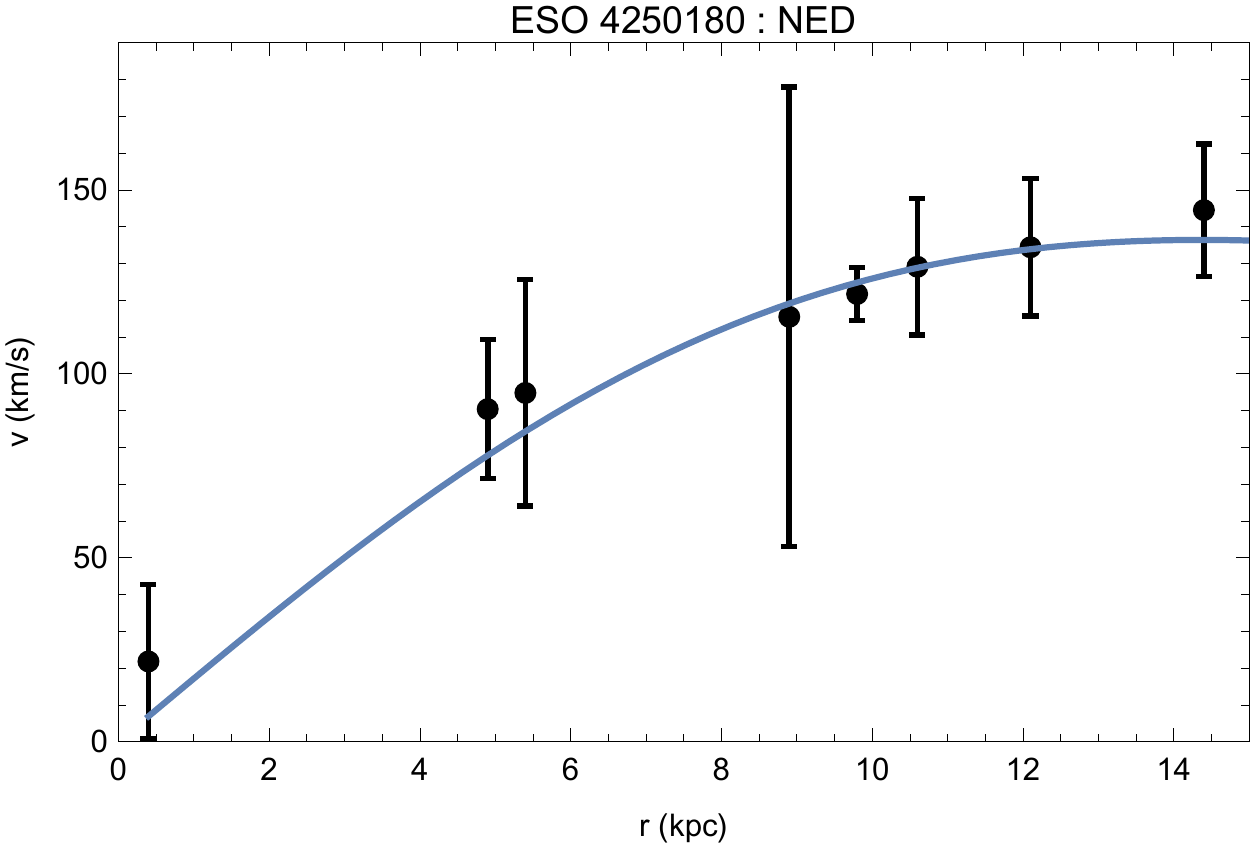} & 
\includegraphics[scale=0.4]{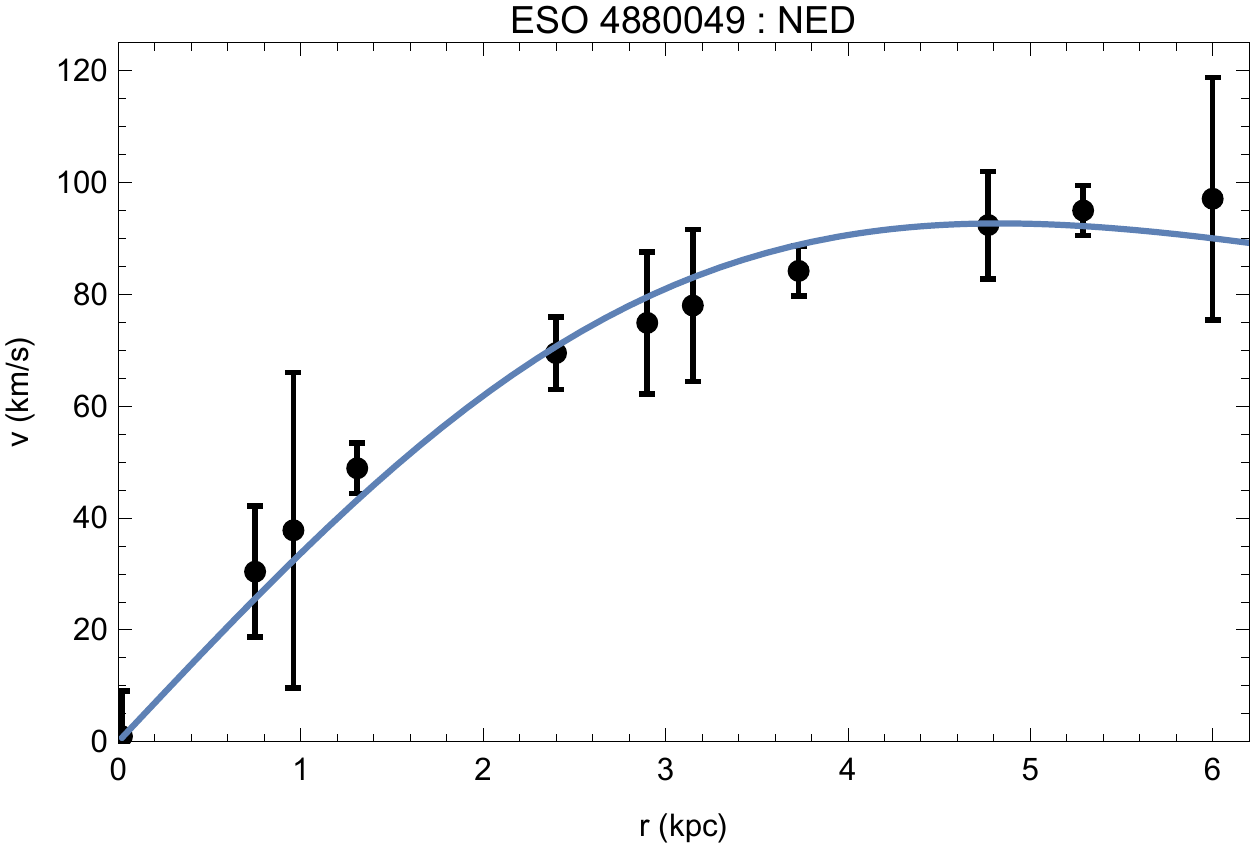} \\ 
\includegraphics[scale=0.4]{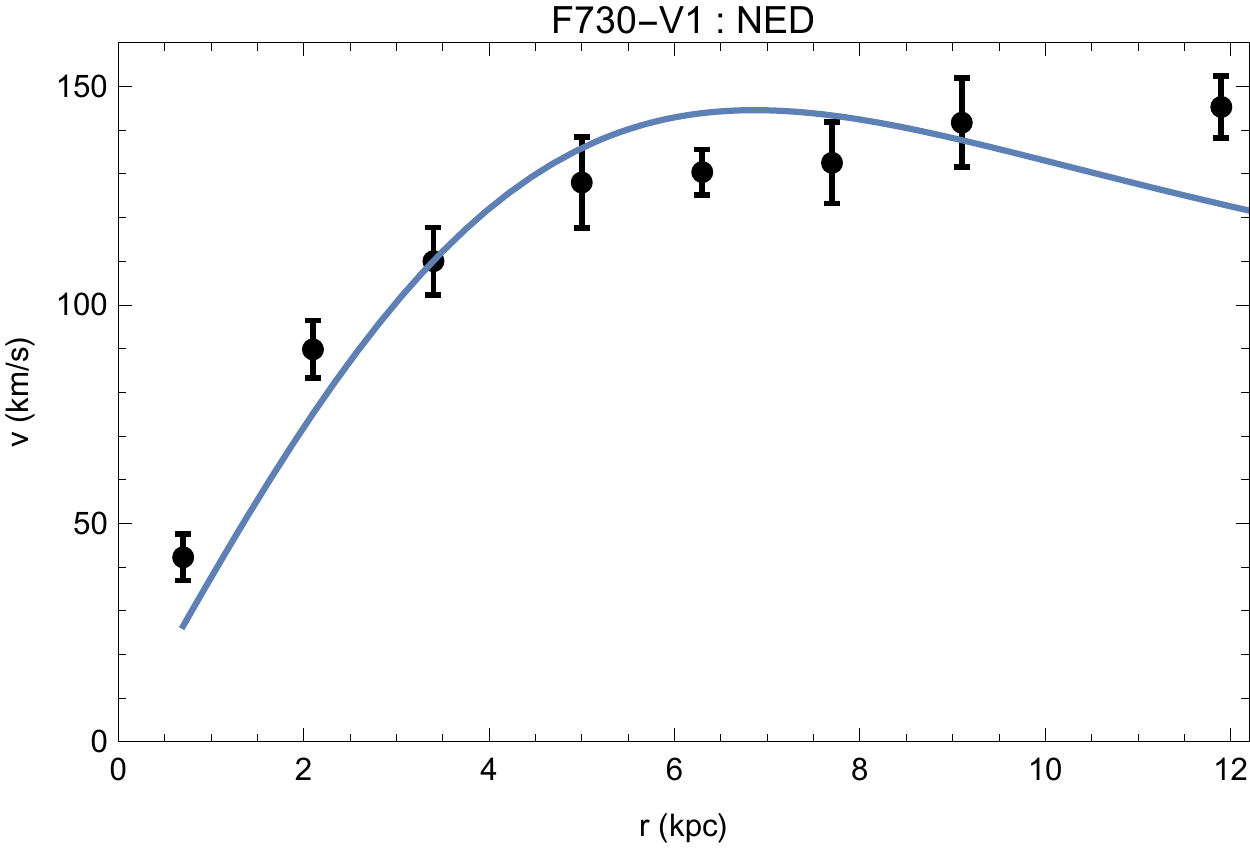} & 
\includegraphics[scale=0.4]{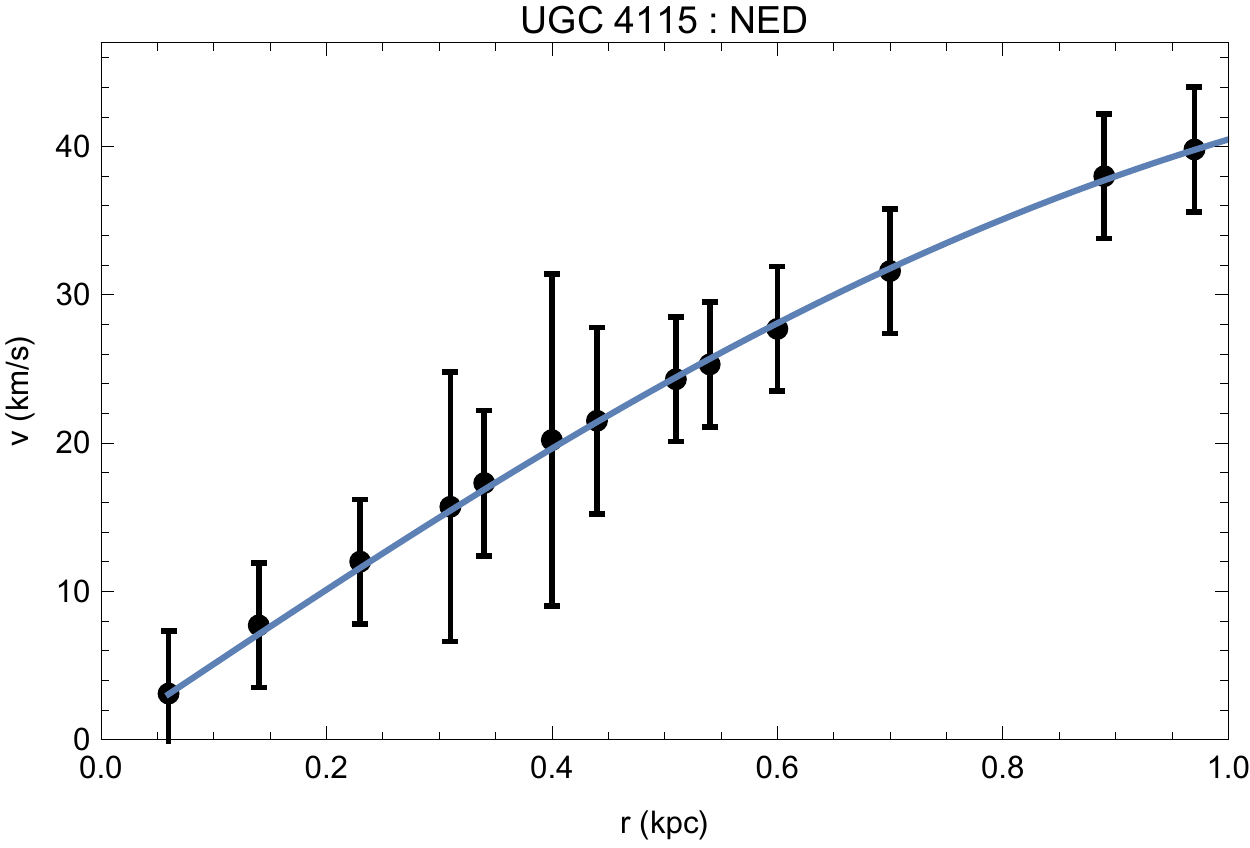} &
\includegraphics[scale=0.4]{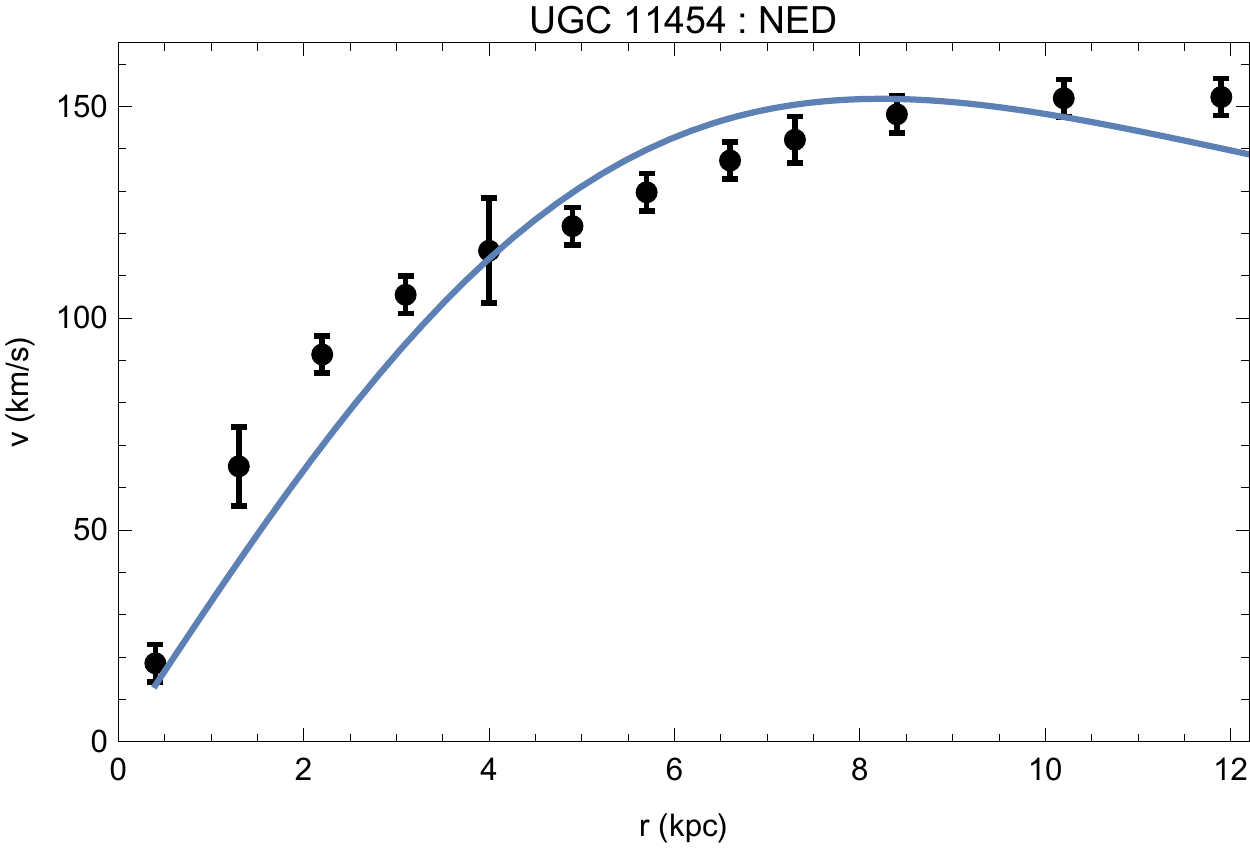} \\ 
\includegraphics[scale=0.4]{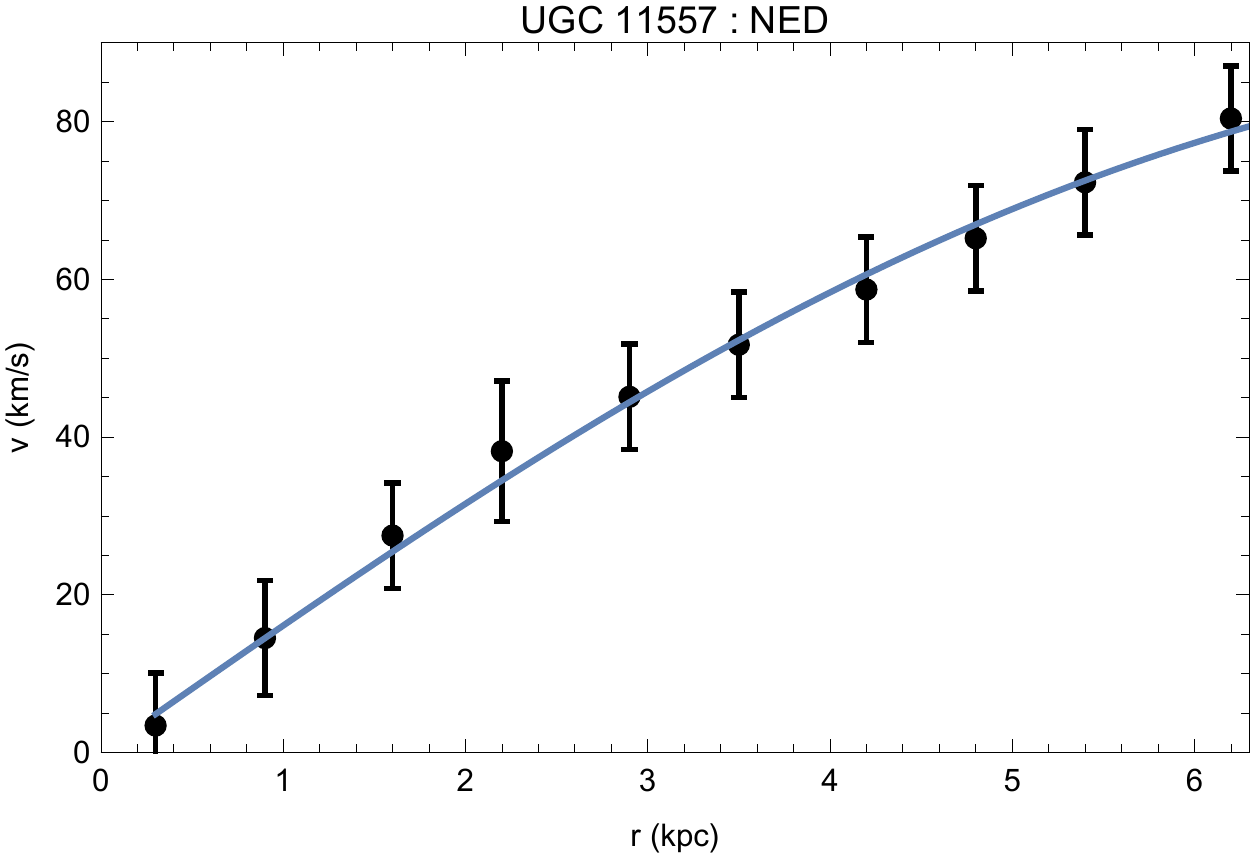} & 
\includegraphics[scale=0.4]{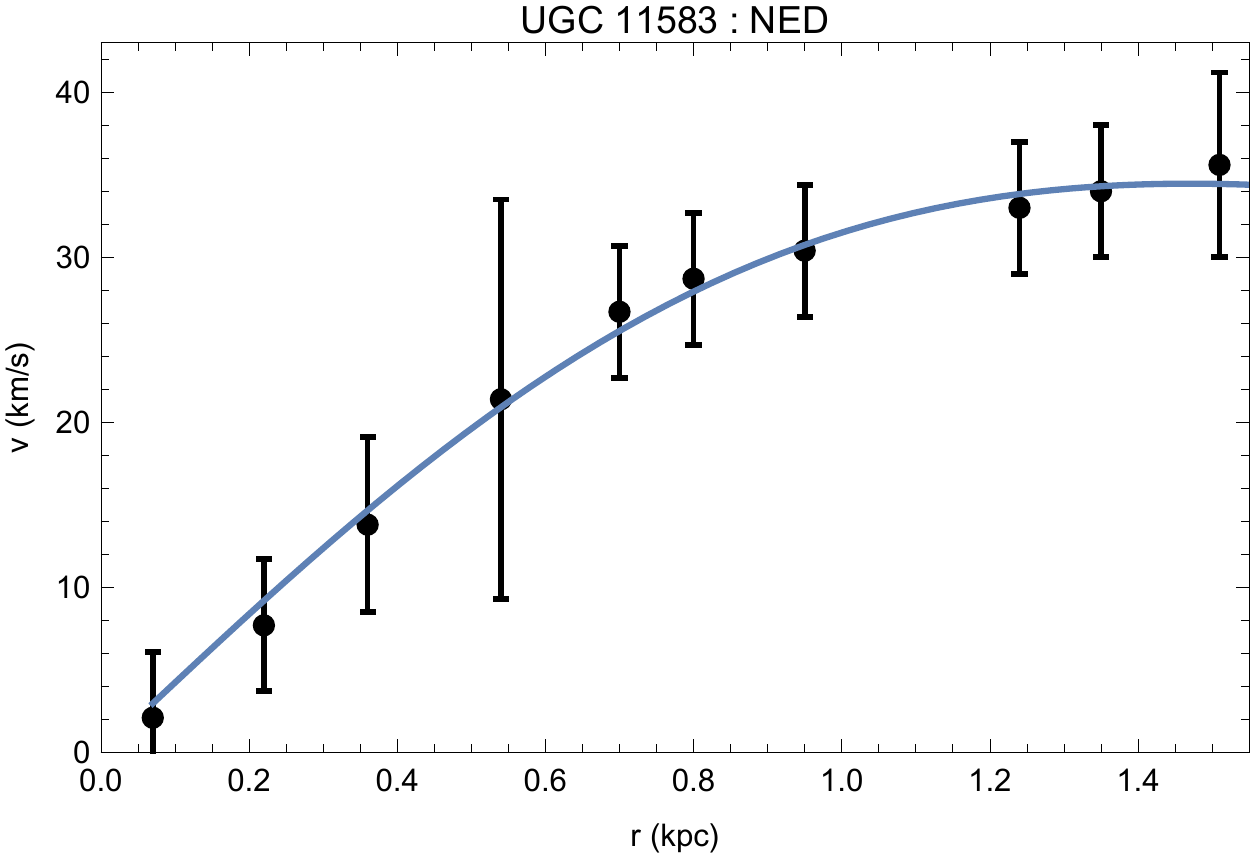} & 
\includegraphics[scale=0.4]{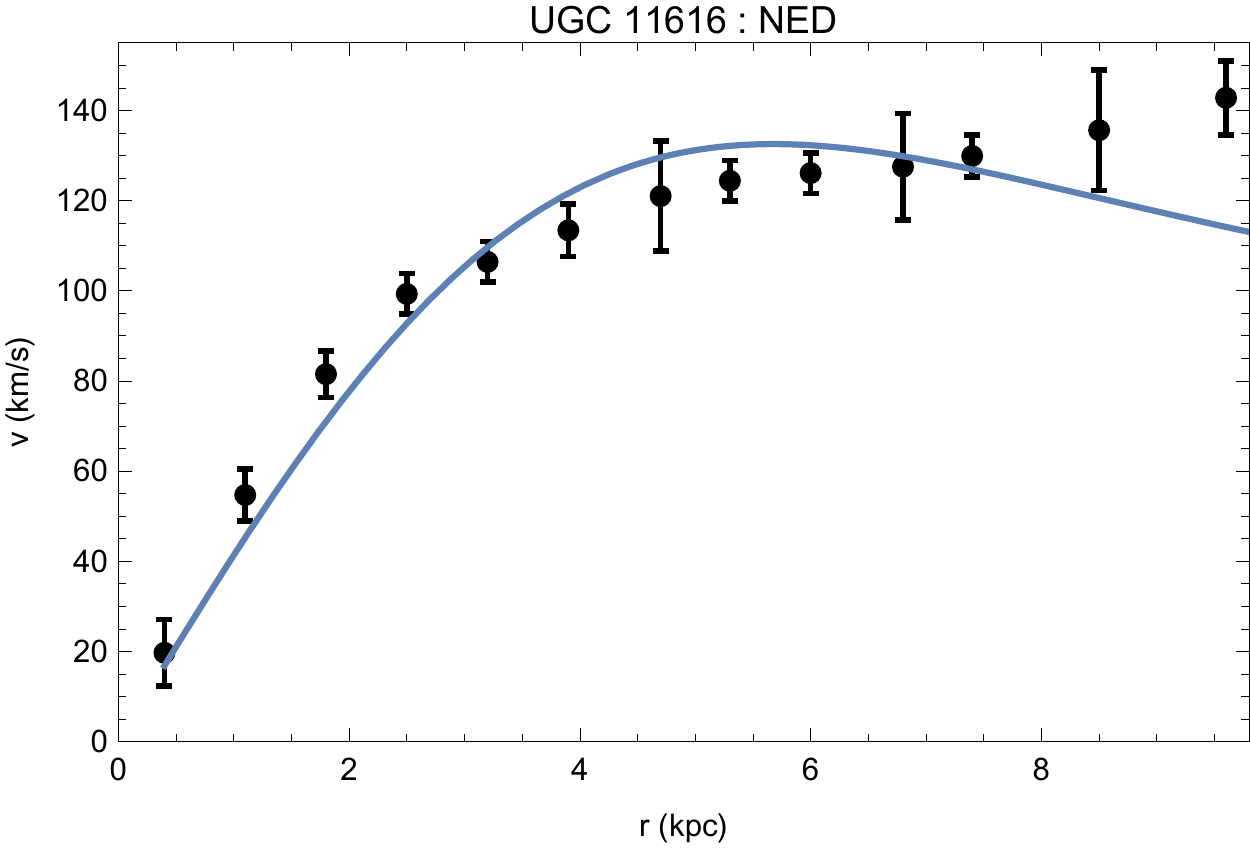} \\  
\includegraphics[scale=0.4]{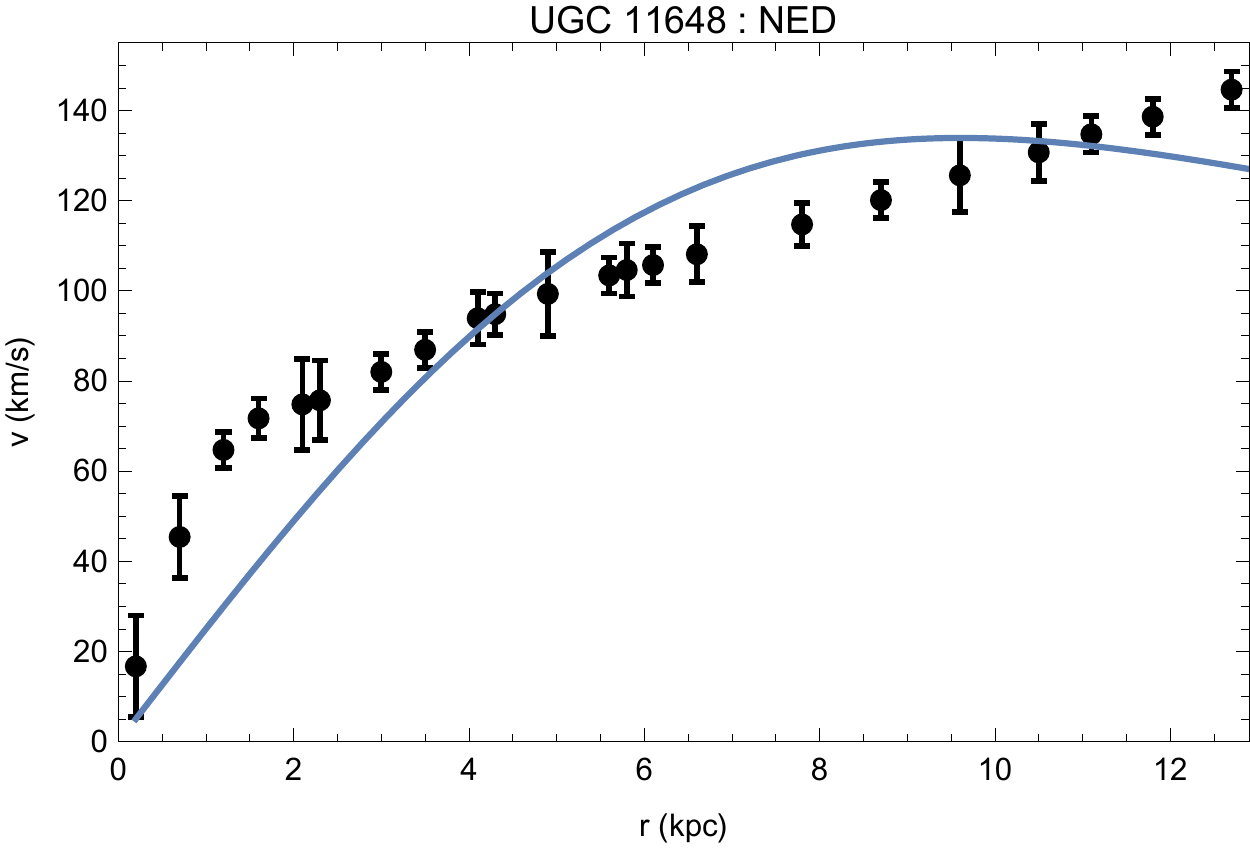} &
\includegraphics[scale=0.4]{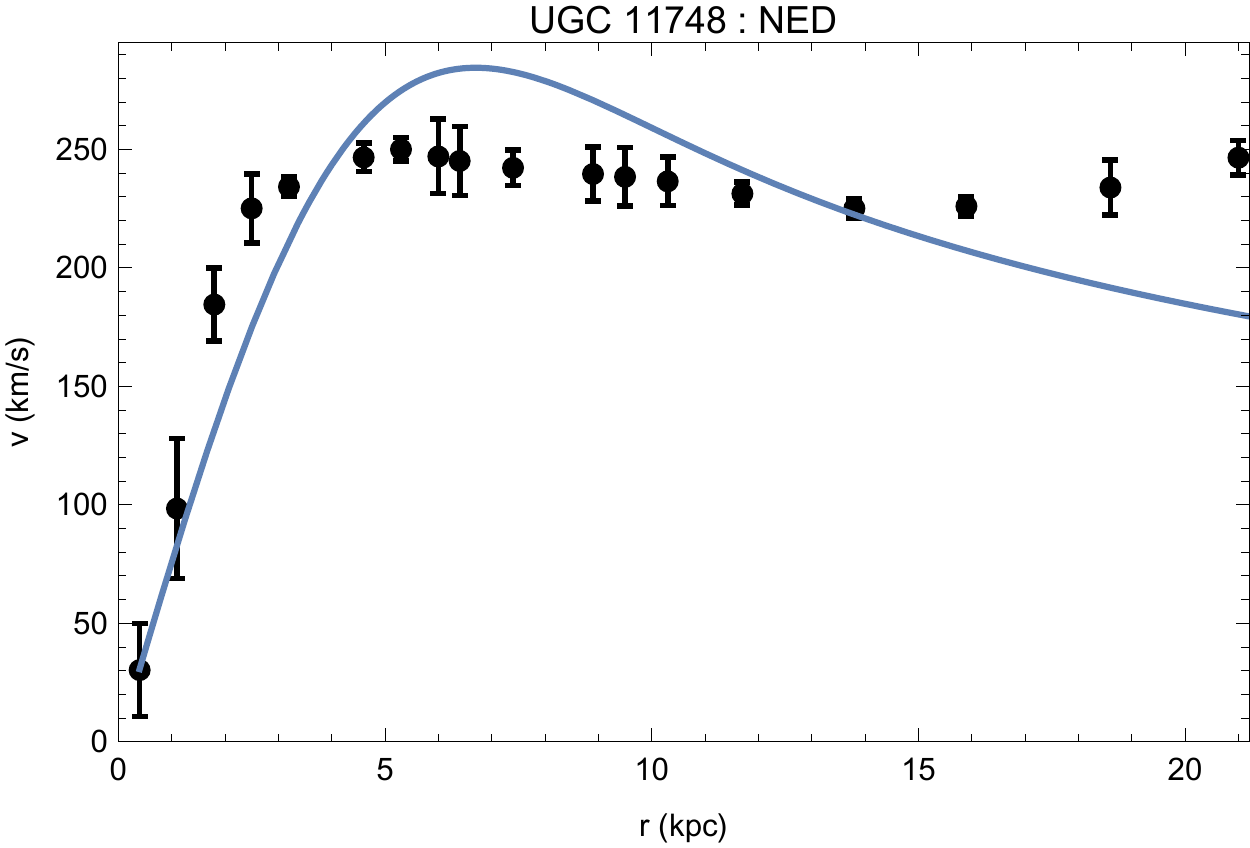} & 
\includegraphics[scale=0.4]{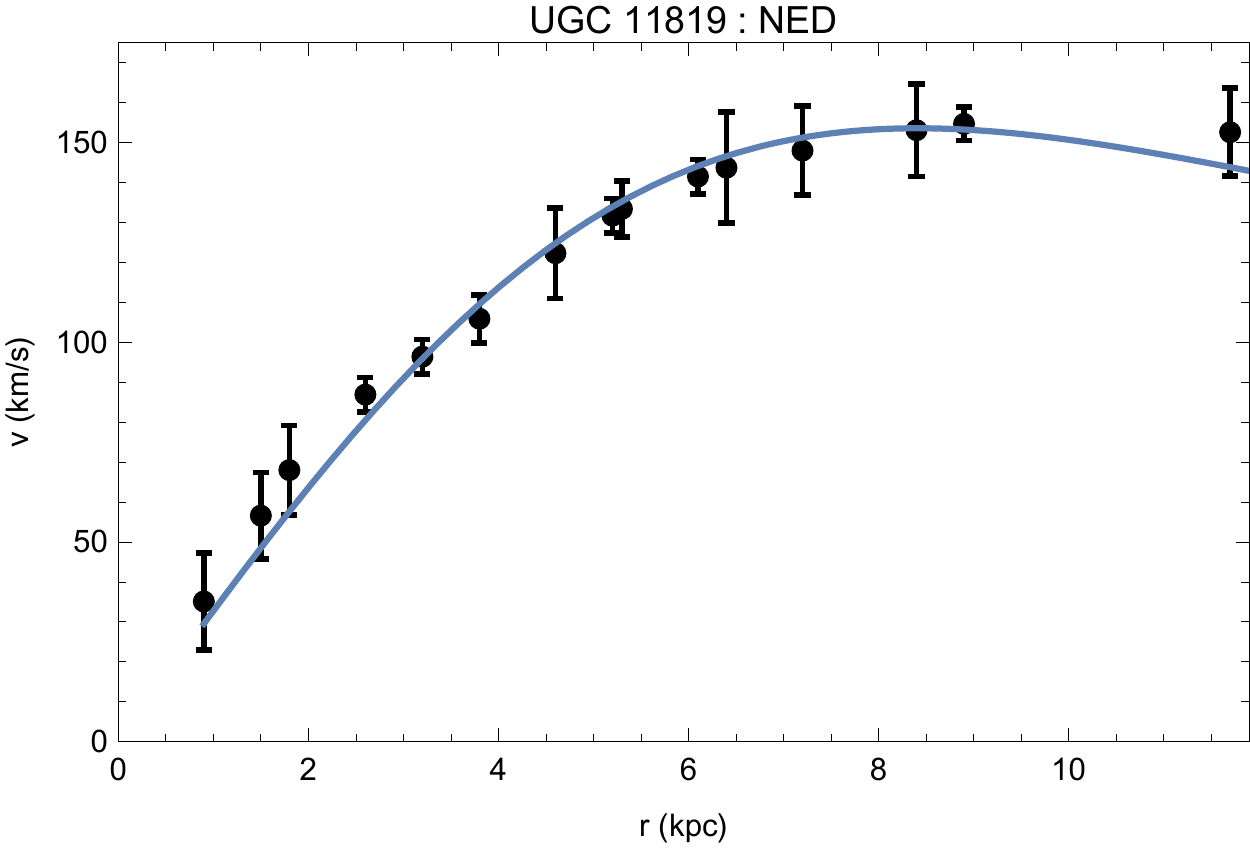}
\end{tabular}
\caption{We show, for each one of the galaxies in the sample in Table \ref{tab:sample}, the plots of the NED theoretical rotation curve (blue solid line), that best fit of the corresponding observational data (black symbols).} 
\label{NC}
\end{figure*}

In Fig. \ref{NC}, it is shown, for each galaxy in the sample of the LSB galaxies, the theoretical fitted curve to a preferred NED value (blue solid line) that best fit to the corresponding observational data (black symbols).
%
%%%%%%%%%%%%%%%%%%%%%%%%%%%%%%%%%%%%%%
%% ROTATION CURVES PDF'S
%%%%%%%%%%%%%%%%%%%%%%%%%%%%%%%%%%%%%%
%
Notice how  galaxies UGC 11648 and UGC 11748 are the worst fitted cases. This was also the case in Ref. \cite{deBlok/etal:2001} for PISO and NFW models. For NED $\chi^2_{\rm red}=11.531$ and $16.596$, respectively.  
Also notice that we have found a preferred range of $\sqrt{\theta}$ values, from $0.5$ to $4.8 \;\rm kpc$, approximately. 

%%%%%%%%%%%%%%%%%%%%%%%%
% Figure muDM for all models
\begin{figure*}[h]
\centering
\begin{tabular}{cc}
\includegraphics[scale=0.65]{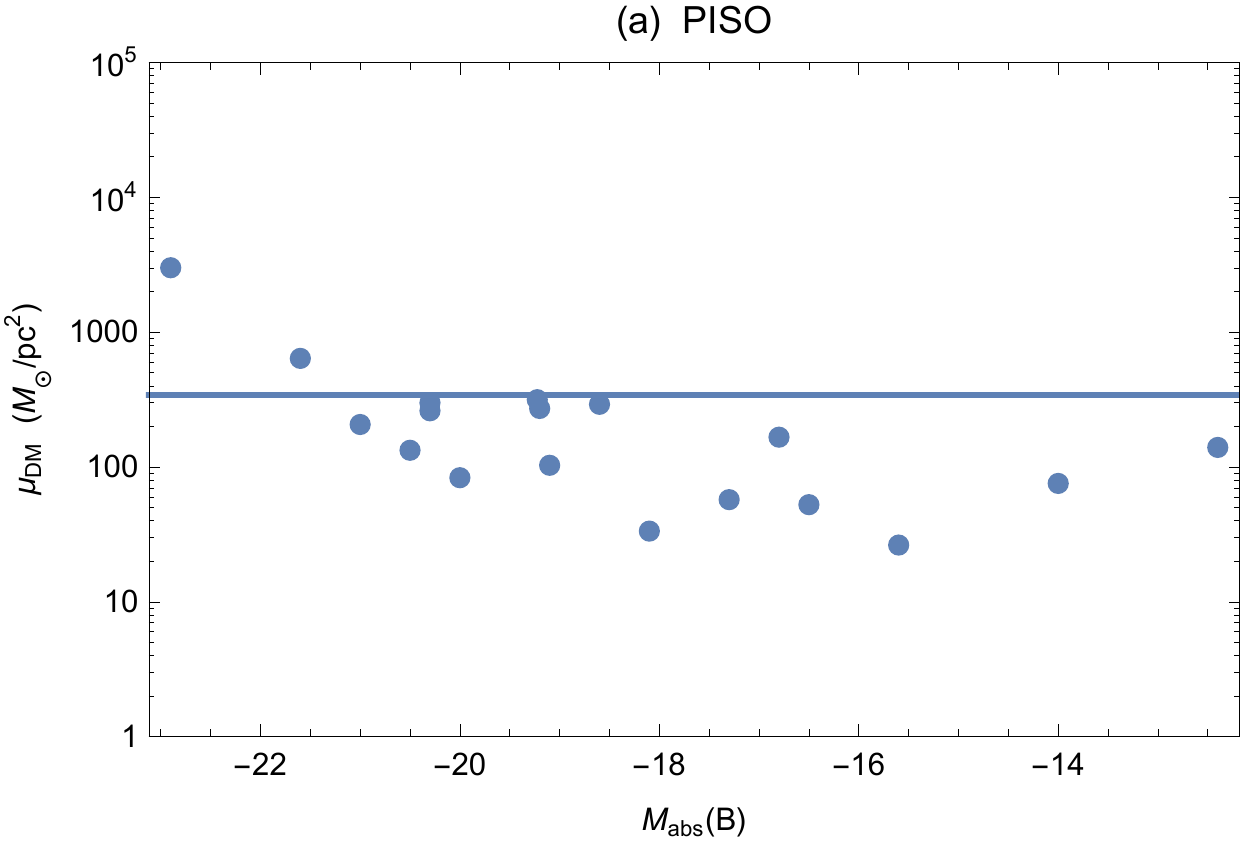} &
\includegraphics[scale=0.65]{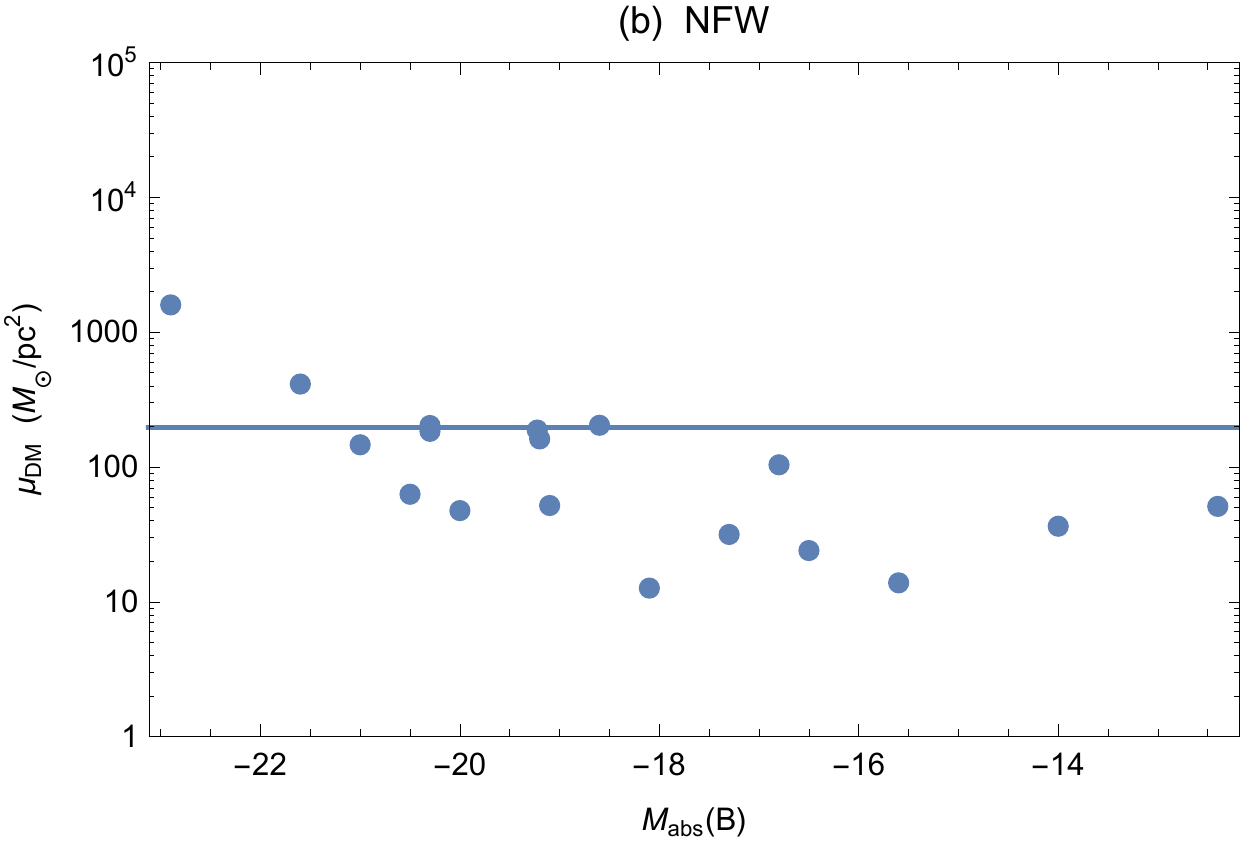} \\ [1\baselineskip]
\includegraphics[scale=0.65]{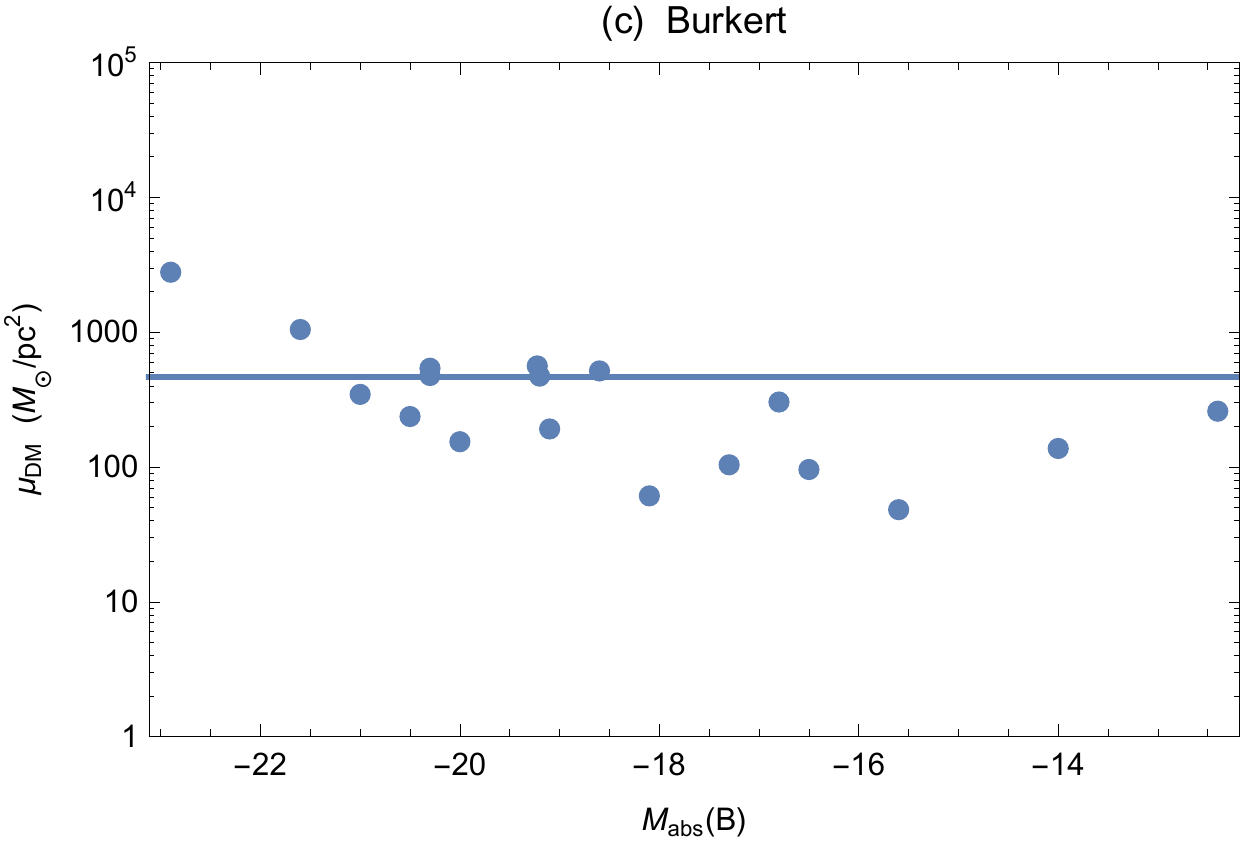} &
\includegraphics[scale=0.65]{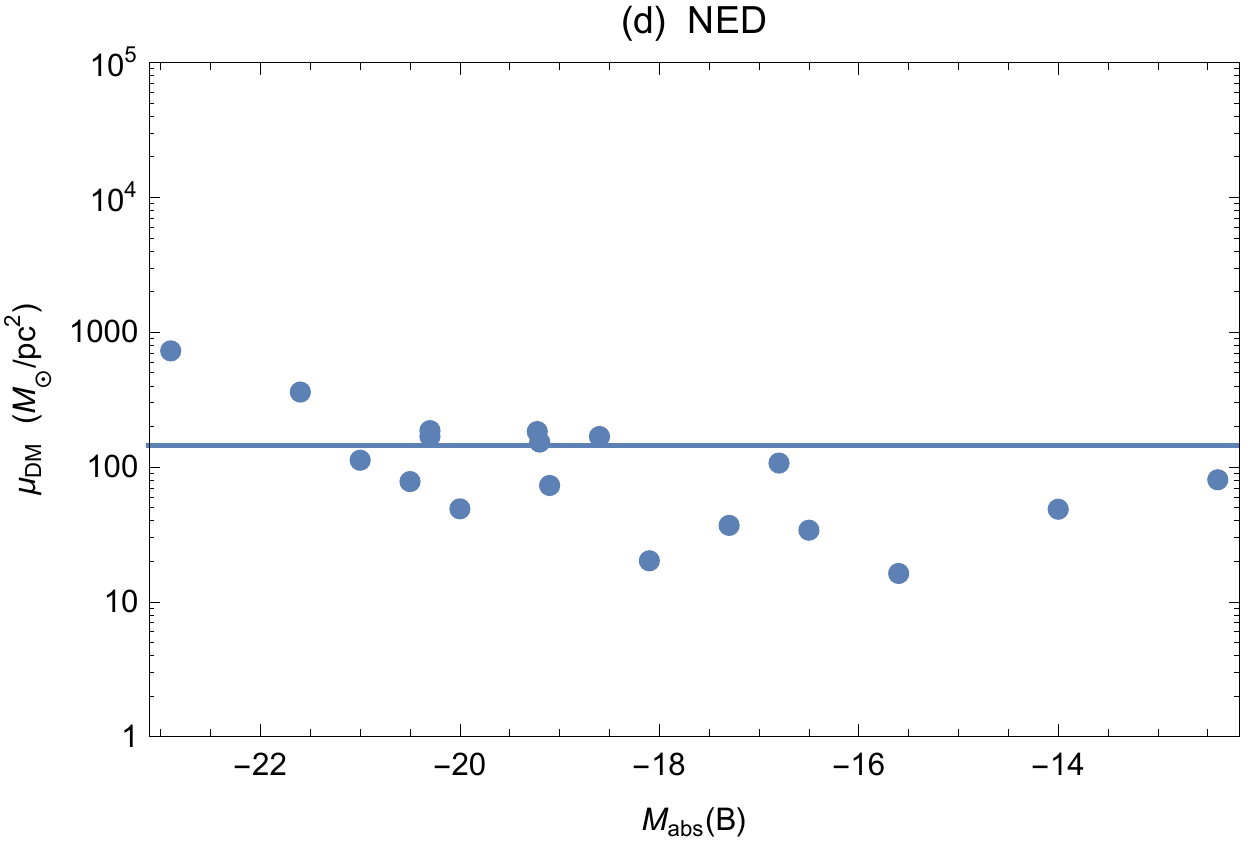} \\ [1\baselineskip]
\includegraphics[scale=0.65]{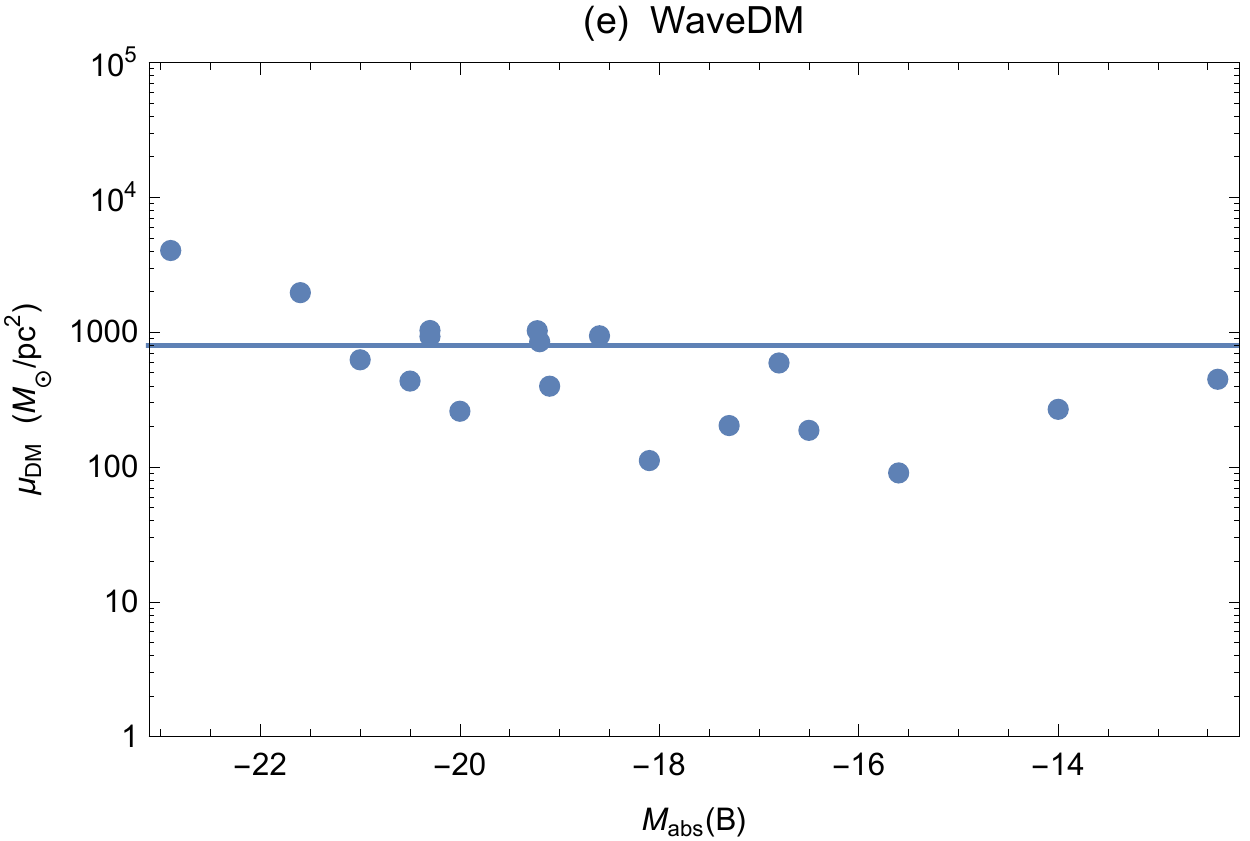} 
\end{tabular}
\caption{Plots of $\mu_{DM}$ versus the absolute magnitude in the B-band of each galaxy (column (5), Table \ref{tab:sample}). 
In (a) is the PISO model; (b) NFW model; (c) Burkert model;(d) NED model and (e) WaveDM model.
The mean value of $\mu_{DM}$ is shown to guide the eye (solid line) in each plot.}\label{fig:muDM}
\end{figure*}
%%%%%%%%%%%%%%%%%%%%%%%%
%
In Ref. 
\cite{2004IAUS..220..377K} was found for PISO model that the surface density $\mu_{DM}$ has nearly a constant value of the order of 100 $M_\odot/$pc$^2$. This work was extended to include more galaxies and other cored DM models in 
\cite{2008MNRAS.383..297S,2009MNRAS.397.1169D}. 
Authors
found that its value is almost constant for galaxies in a width range of magnitude values. This indicates that  
$\mu_{DM}$ should be a quantity characterizing dark matter in galaxies. Following the results in Ref. \cite{2016PhRvL.117t1101M} that used the newly SPARC galaxy catalog, a theoretical study was done in Ref. 
\cite{2017arXiv170205103U}. We redo these theoretical estimations to find the theoretical values for the DM models
we are interested in: PISO, NFW, Burkert, NED and WaveDM. 
There were differences among PISO and Burkert values we have obtained with the values reported in Ref. \cite{2017arXiv170205103U}. 
In Tables \ref{tab:PISO} for PISO model, \ref{tab:NFW} for NFW model, \ref{tab:Burkert} for Burkert model, \ref{tab:NED} for NED model and \ref{tab:WDM} for WaveDM model, we reported this surface density values, column (4) in each table. In Fig. \ref{fig:muDM} we have plotted $\mu_{DM}$ versus $M_{\text{abs}}(B)$ for all models.

For PISO model the mean value of $\mu_{DM}$ is $341.64\, M_\odot/$pc$^2$ while the theoretical value we found, following 
\cite{2017arXiv170205103U}, is $193\, M_\odot/$pc$^2$. For NFW: $\mu_{DM}$ is $195.59\, M_\odot/$pc$^2$ versus the theoretical value of $89\, M_\odot/$pc$^2$. 
For Burkert: $\mu_{DM}$ is $462.11\, M_\odot/$pc$^2$ versus the theoretical value of $348.69\, M_\odot/$pc$^2$. 
For NED DM model we have that the mean value of $\mu_{DM}$ is $144.21\, M_\odot/$pc$^2$ versus its theoretical value of $\mu_{DM}$ is $116.97\, M_\odot/$pc$^2$. 
And for WaveDM model we obtained that the mean value of $\mu_{DM}$ is $797.59\, M_\odot/$pc$^2$
while the theoretical value is $648.58\, M_\odot/$pc$^2$.
This work should be extended to consider more galaxies in order to test more throughly the theoretical predictions as it was done in Ref. \cite{2009MNRAS.397.1169D}.

%%%%%%%%%%%%%%%%%%%%%%%%
% Figure masses of the DM for all models
\begin{figure*}[h]
\centering
\begin{tabular}{cc}
\includegraphics[scale=0.65]{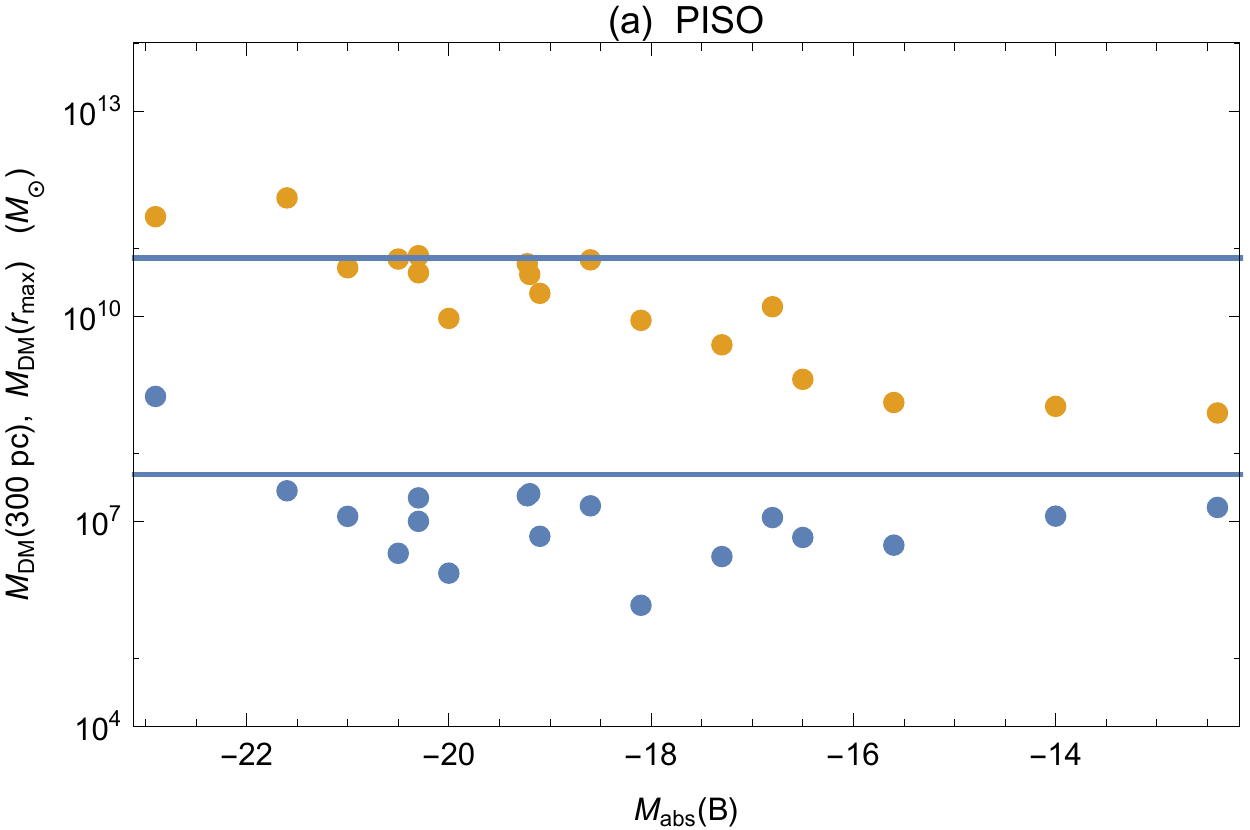} &
\includegraphics[scale=0.65]{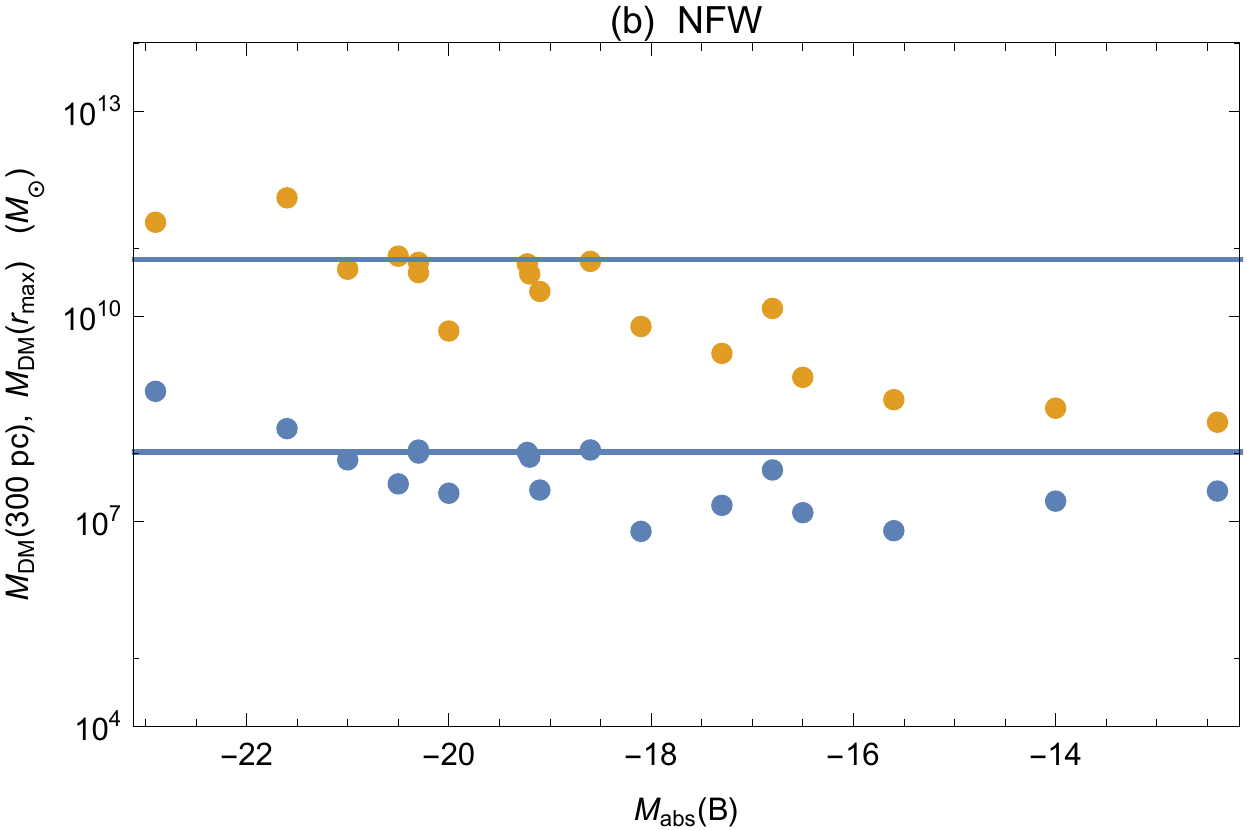} \\ [1\baselineskip]
\includegraphics[scale=0.65]{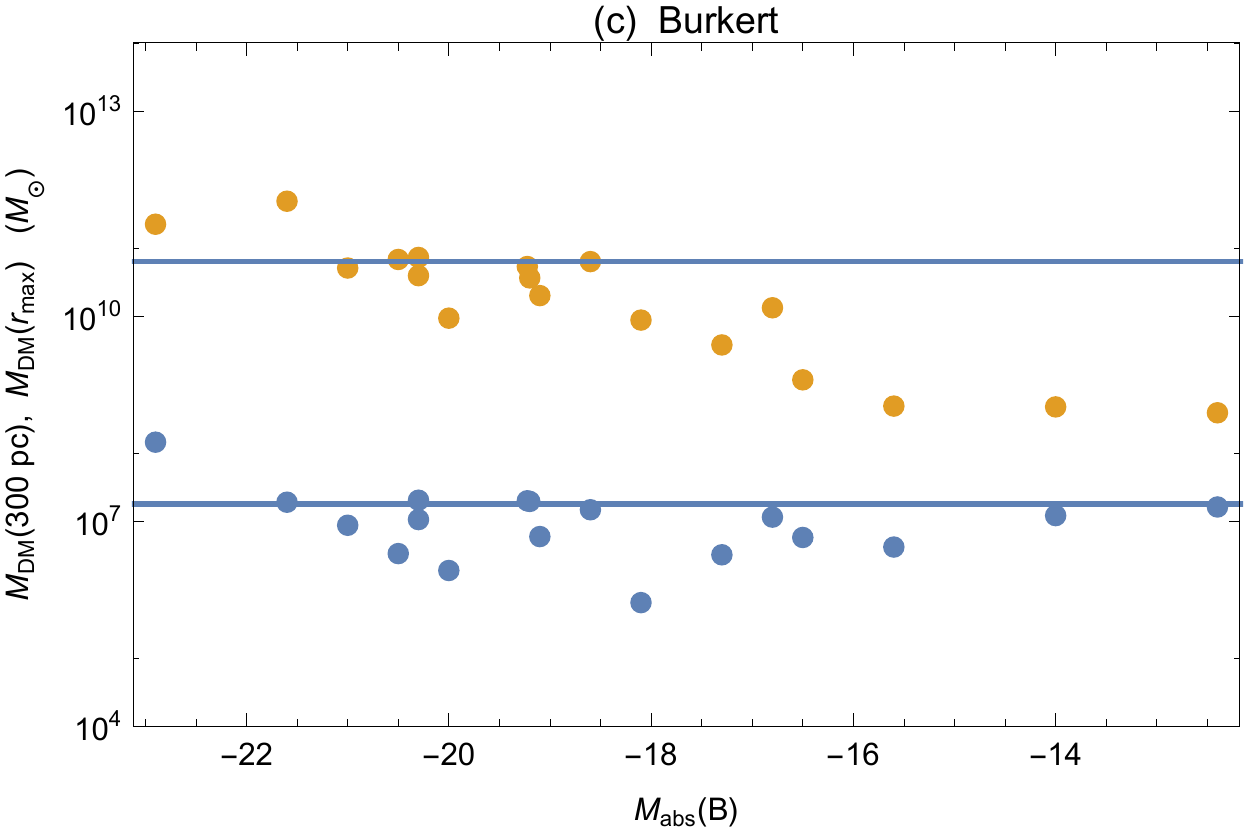} &
\includegraphics[scale=0.65]{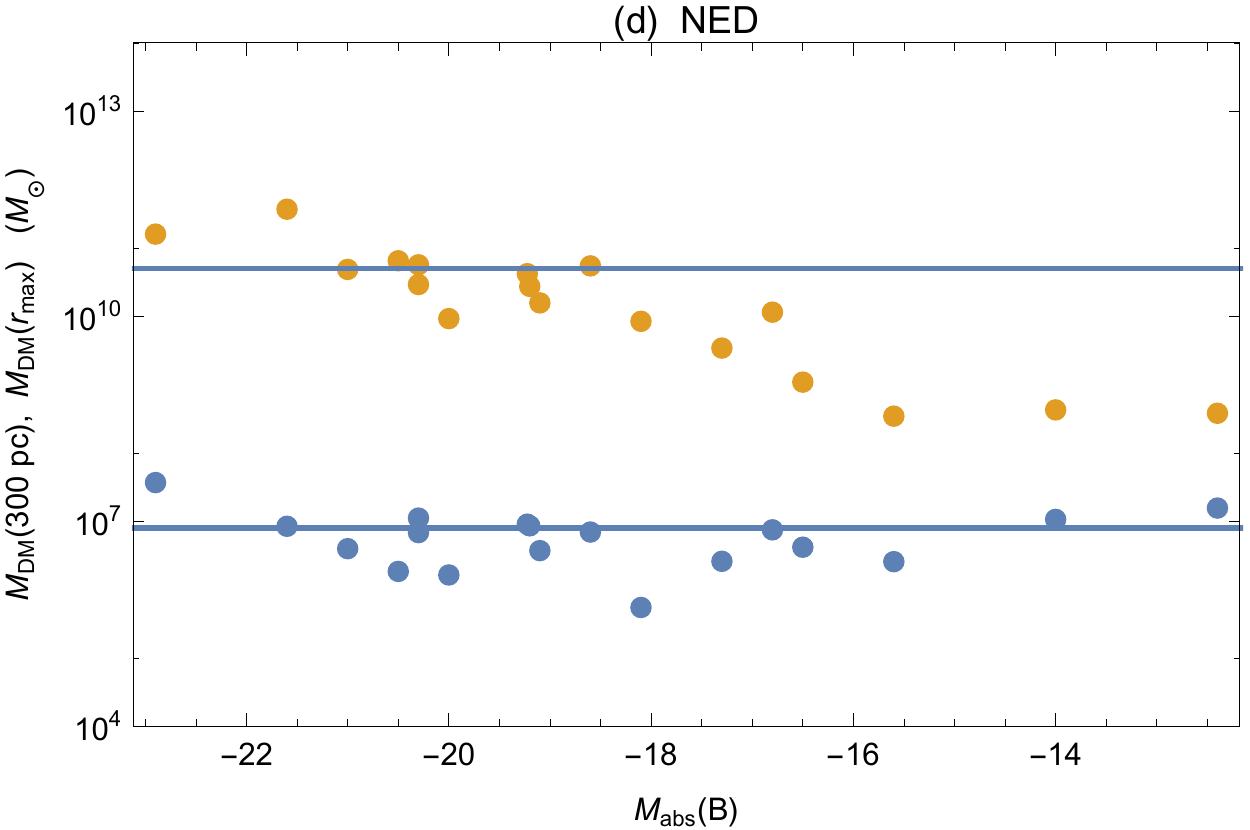} \\ [1\baselineskip]
\includegraphics[scale=0.65]{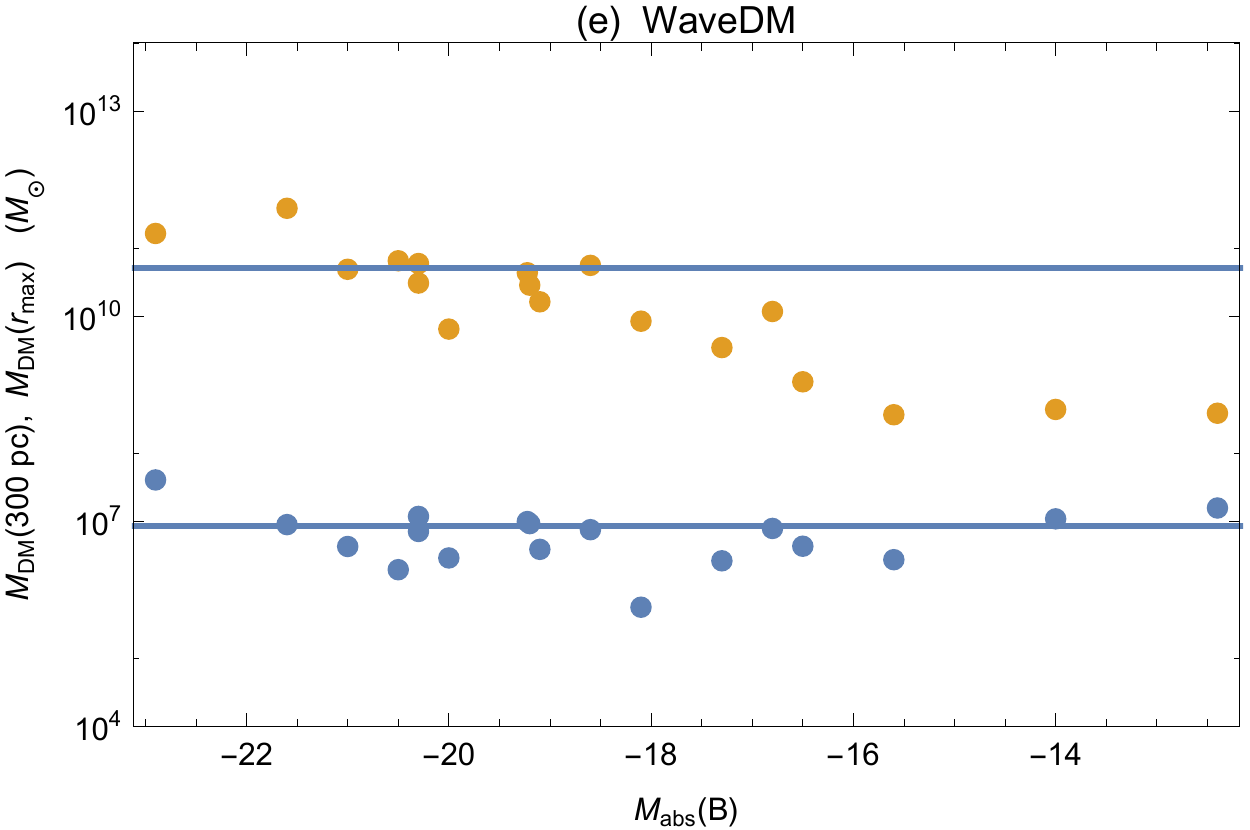} &
\end{tabular}
\caption{Plots of the $M_{DM}$ versus the absolute magnitude in the B-band of each galaxy (column (5), Table \ref{tab:sample}).
In (a) is the PISO model; (b) NFW model; (c) Burkert model; (d) NED model and (e) WaveDM model.
The mean value of $M_{DM}(300$ pc$)$ and $M_{DM}(r_{\text{max}})$ is shown to guide the eye (solid line).}\label{fig:masses}
\end{figure*}
%%%%%%%%%%%%%%%%%%%%%%%%
%

In column (5) of each one of the Tables \ref{tab:PISO}, \ref{tab:NFW}, \ref{tab:Burkert} and \ref{tab:NED}, we have reported values of the masses of the DM models up to $300$ pc, $M_{DM} (300$ pc$)$, and they were plotted versus $M_{\text{abs}}(B)$ (lower filled circles) for all models in Fig. \ref {fig:masses}. To guide the eye we have plotted the mean values of 
$M_{DM} (300$ pc$)$.
The mean values we found for each one of the models are: 
PISO: $4.85\times 10^{7} \, M_\odot$;
NFW: $1.03\times 10^{8} \, M_\odot$;
Burkert: $1.79\times 10^{7} \, M_\odot$;
NED: $0.79\times 10^{7} \, M_\odot$;
WaveDM: $0.85\times 10^{7} \, M_\odot$.
In Ref. \cite{2008Natur.454.1096S} was reported this mass within the central $300$ pc for 18 dSph Milky Way satellites. The value found is about $10^7 \, M_\odot$ which is consistent with the mean values we found except for NFW DM model whose value is an order of magnitude larger. This would indicate that NFW is a model that can not explain rotation curves in spiral galaxies.

Also, in Tables \ref{tab:PISO}, \ref{tab:NFW}, \ref{tab:Burkert}, \ref{tab:NED} and \ref{tab:WDM}, column (6), we reported values for the total mass of the DM models up to the last spatial point measured, $r_\text{max}$. This mass, $M_{DM}(r_\text{max})$, was plotted in Fig. \ref {fig:masses} (upper filled circles). Values of this total mass fail to follow the constant mean value, almost of the order of $10^{10} \, M_\odot$ for all models.
 In the case of NED, this is in tension with the result reported in Ref. \cite{Hernandez-Almada:2016ioe}. But we believe more work is needed in order to make a final conclusion.

\begin{figure*}[h]
\centering
\includegraphics[scale=0.5]{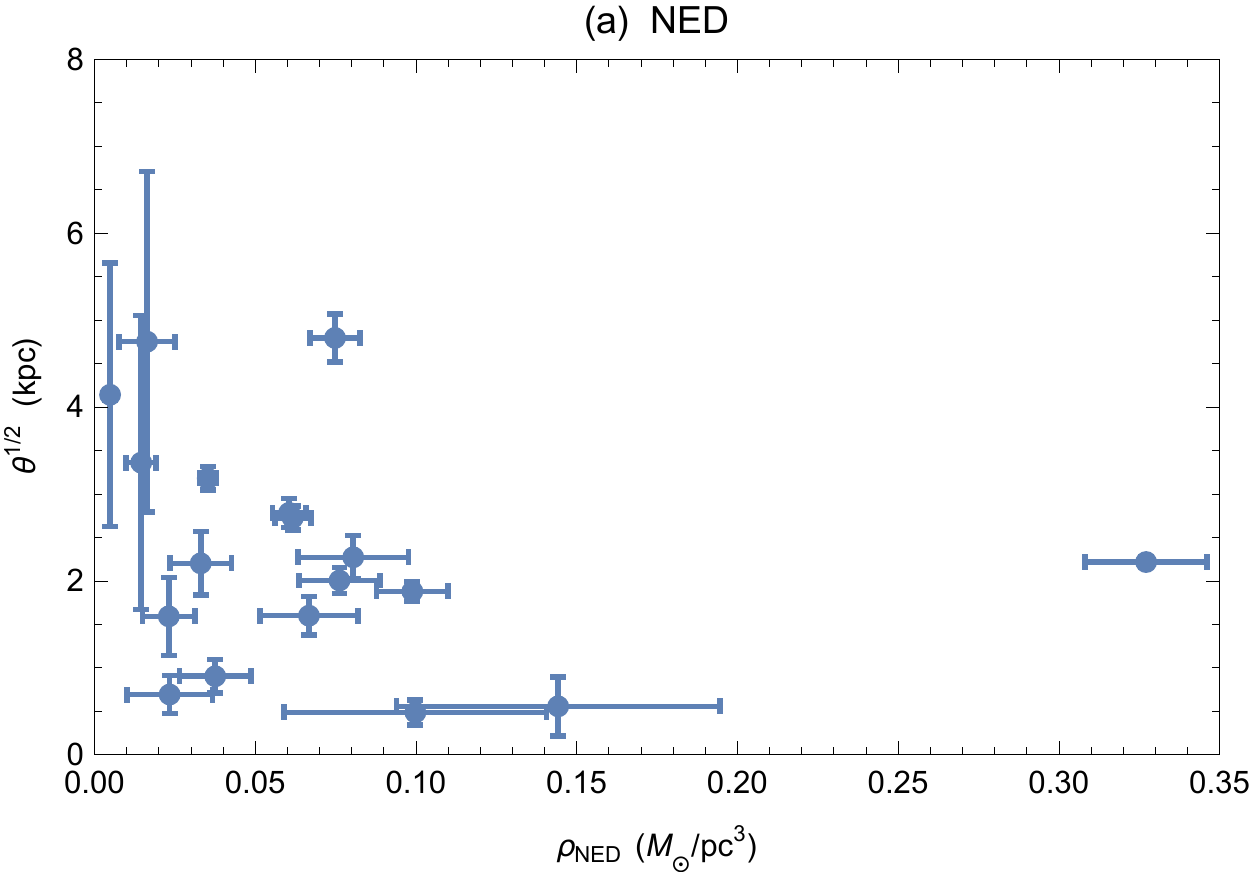} \\ [1\baselineskip]
\includegraphics[scale=0.5]{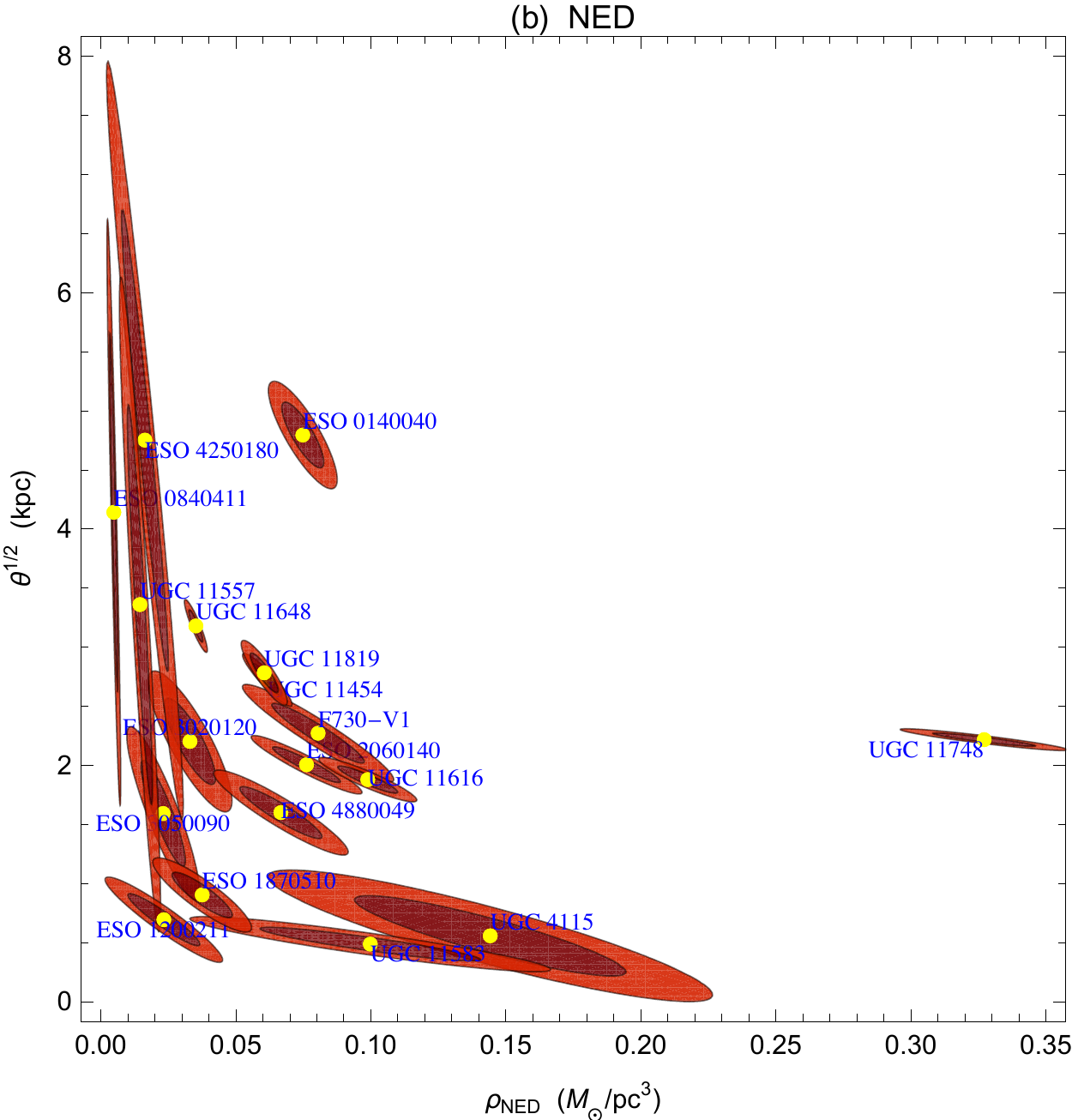} 
\caption{(a) Plots of $\sqrt\theta$ versus $\rho_{NED}$ for the NED model, errors bars are also shown. 
In (b) is the plot of contours at $1\sigma$ and $2\sigma$ of the estimated parameters for each one of the rotation curves of the galaxies studied in this work.}\label{fig:countours}
\end{figure*}

Finally, for NED model, Fig. \ref{fig:countours} shows in (a) the fitted values of parameters $\sqrt\theta$ versus 
$\rho_{NED}$ that include the error bars. In (b) it is shown contour plots for each galaxy at $1\sigma$ and $2\sigma$. This latter plot shows that $\sqrt\theta$  and $\rho_{NED}$ are anti-correlated as is always the case for this kind of rotation curve models described by two parameters in a general form similar to Eq. \eqref{velrotnonadim}.

%%%%%%%%%%%%%%%%%%%%%%%%
% Figure masses of the DM for all models
%%%%%%%%%%%%%%%%%%%%%%%%
%
%%%%%%%%%%%%%%%%%%%%%%%%%%%%%%%%%%%%%%%%
\section{Discussion and conclusions} \label{CR}
%%%%%%%%%%%%%%%%%%%%%%%%%%%%%%%%%%%%%%%%%

In this paper we have assumed that NED is an important component in Universe dynamics affecting the stellar equations of motion and the dynamic of galaxies in special the rotation curves. From here, it is possible to discuss the research shown in this paper in two main cases:

In the first case, we use the dynamic equations for the evolution of stars with the aim of investigate the behavior dictated by the presence of an uniform density (incompressible fluid) and NED. We find a constraint for the parameter $\sqrt{\theta}$ under the premise of observing important effects in the dynamics of effective pressure and mass, always comparing with the traditional knowledge of the uniform density stars. In this vein, we give to the task of constraint the NED parameter as: $\sqrt{\theta}<0.753R$, emphasizing how the effects induced by NED, generates a not expected behavior.

In addition, with polytropic stars it is possible to model Dwarf Stars with index $n=3$. Here we also add NED which does not interact with the polytropic fluid, presenting a modified Lane-Emden equation to describe the dynamics of the star, our results present important effects in the dynamics except for values 
that fulfill the conditions 
$\kappa\geqslant10^2\sqrt{\theta}$ and $\kappa\leqslant\sqrt{\theta}$. We establish our bounds, with a plot in which the correction is negligible, that is, when $\bar{\theta}\to0$, hence, we can infer those that are mimicked with the traditional limit. 

The second case took into account the corresponding rotation velocity of galaxies, assuming that NED is related with the galaxy halo. From here, we compute the rotation velocity associated with this model and was compared with four of the most studied and accepted models in literature, which are: PISO, NFW, Burkert and WaveDM. 
%{\color{blue}
The analysis was implemented through a $\chi^2$ best fit, with eighteen high resolution rotation curves of LSB galaxies with no photometry. 
The range of values predicted by the method was from $\simeq 0.5$  kpc to $\simeq 5$  kpc. 
Also, the $\chi^2$ best fit for NED was not the most accurate in comparison with the other standard models (see Tables \ref{tab:PISO}, \ref{tab:NFW} and \ref{tab:Burkert}); 
however it is important to remark that although it was not the best fitting model, the advantage lies in the comparison with NC which is inspired by the internal property of the space-time and not relinquish to this characteristics.
%}

For each DM model and for each galaxy in the analyzed sample we computed three important quantities that should give us information about galaxy formation and evolution that are summarized in Table \ref{tab:muM300}: 
the surface density $\mu_{DM}$ and the masses within the 300 pc, $M_{DM}(300$ pc$)$, and $M_{DM}(r_\text{max})$ for the DM haloes. From other studies it was expected that these quantities were nearly constants, independent of the absolute magnitude of the galaxies\cite{2004IAUS..220..377K,2017arXiv170205103U,2008Natur.454.1096S,Hernandez-Almada:2016ioe}. 
The general tendency, found in this work, is that these quantities are roughly constant (except the behavior of  $M_{DM}(r_\text{max})$).
We found like in \cite{2008Natur.454.1096S}, for dwarf spheroidal galaxies, that also must be a common mass for spiral galaxies within $300$ pc and with the same order of magnitud, $10^{7}$ $M_\odot$. This would give a central density for DM of $\sim 0.1 M_\odot$pc$^{-3}$, independent of the DM model 
with the exception of NFW model whose value is an order of magnitud larger. If this result is confirmed NFW would have problems to explain the observed rotation curves of spiral galaxies.
The theoretical and observational predictions give that the surface density, $\mu_{DM}$, is constant, independent of the absolute magnitud of the galaxy, but it depends on the DM model. We also found a roughly constant behavior in the surface density in each one of the analyzed models. On the other hand the total mass up to the outer spatial point, $r_\text{max}$ fail to have this behavior, this is in tension with a previous result found in Ref. \cite{Hernandez-Almada:2016ioe}.
However, we believe that it is needed a more systematic study with a better sample of galaxies to reach a final conclusion. 

Our results about NED parameter are summarized in Table \ref{tab:final} showing the difference between each constraint. It is possible to observe that stellar dynamics provides a lower bound for the free parameter meanwhile galaxy rotation curves provides a higher bounds: an interval of values for the NED parameter. Nevertheless, the rotation curves analysis is less restrictive unlike the toy model cases of stellar analysis; presenting the first one a better confidence in the value. We emphasize that it is necessary to perform a further analysis in both cases to uncover and narrow the free parameter for this \emph{matter} inspired by NED models. 
Each model provide us with different results, which are subjected to the astrophysical system to which it is being addressed. However, it is possible to establish a range of values where the NED parameter must fit in.

%%%%%%%%%%%%%%%%%%%%%%%%
% Compilation of mu and mass
\begin{center}
\begin{table*}
\ra{1.3}
\begin{center}
\begin{tabular}{@{}l l r r r r r r r r r r r @ {}}\toprule
    \hline \hline
    \multicolumn{12}{c}{Results Compilation for $\mu_{DM}$ and $M_{DM}(300$ pc$)$} \\
    \hline 
  %  \multicolumn{12}{c}{Observational data}\\
   %\hline 
\multicolumn{1}{c}{} &
\multicolumn{1}{c}{PISO }&
\multicolumn{1}{c}{NFW} &
\multicolumn{1}{c}{Burkert}&
\multicolumn{1}{c}{NED} &
\multicolumn{1}{c}{WaveDM}
%\multicolumn{1}{c}{$\rho(0)$ (MeV$^4$) } &
%\multicolumn{1}{c}{$\sqrt{\theta}$($10^{18}$MeV$^{-1}$$\geqslant/\leqslant10^{-2}$)} 

\\
 
\cmidrule[0.4pt](r{0.125em}){1-1}
\cmidrule[0.4pt](lr{0.125em}){2-2}
\cmidrule[0.4pt](lr{0.125em}){3-3}
\cmidrule[0.4pt](lr{0.125em}){4-4}
\cmidrule[0.4pt](lr{0.125em}){5-5}
\cmidrule[0.4pt](lr{0.125em}){6-6}
\cmidrule[0.4pt](lr{0.125em}){7-7}
\cmidrule[0.4pt](lr{0.125em}){8-8}
\cmidrule[0.4pt](lr{0.125em}){9-9}
\cmidrule[0.4pt](lr{0.125em}){10-10}
\cmidrule[0.4pt](lr{0.125em}){11-11}
\cmidrule[0.4pt](lr{0.125em}){12-12}
$\hat{r}$ & 1.52 & 0 & 0.96  & 1.94 & 0.36 \\
$\hat{g}_\text{max}$ & 2.89 & $2\pi$ & 1.60  & 4.77 & 0.86 \\
$\langle\mu_{DM}\rangle$ & 341.64 & 195.59 & 462.11 & 144.21 & 797.59 \\ 
$\mu_{DM}$ & 193.00 & 88.73 & 348.69 & 116.97 & 648.58\\ 
$\langle M_{DM} (300 $ pc$) \rangle$ & $4.85 $ & 
$10.3 $ & $1.79 $ & $0.79 $ & 0.85 \\ 

 \bottomrule
\hline \hline
\end{tabular}
\end{center}
\caption{Compilation of the results 
for $\mu_{DM}$ and $M_{DM}(300$ pc$)$
shown through the paper. Here $\hat{r}=r/r_i$, where the theoretical maximum is found for 
the dimensionless acceleration
$\hat{g}$ \cite{2017arXiv170205103U}, $r_i$ is the scale length for each model, and 
$\langle\mu_{DM}\rangle$ and $\langle M_{DM} (300 $ pc$) \rangle$ 
 are the mean value of $\mu_{DM}$ and $M_{DM} (300 $ pc$)$, respectively, 
 we found for each galaxy in the analyzed sample. 
 $\langle\mu_{DM}\rangle$ and $\mu_{DM}$ are in $M_\odot$/pc$^2$ and $M_{DM}(300$ pc$)$ is in 
 $10^7 M_\odot$.
 }\label{tab:muM300}
\end{table*}
\end{center}
%%%%%%%%%%%%%%%%%%%%%%%%

%%%%%%%%%%%%%%%%%%%%%%%%
% Compilation
\begin{center}
\begin{table*}
\ra{1.3}
\begin{center}
\begin{tabular}{@{}l l r r r r r r r r r r r @ {}}\toprule
    \hline \hline
    \multicolumn{12}{c}{Results Compilation} \\
    \hline 
  %  \multicolumn{12}{c}{Observational data}\\
   %\hline 
\multicolumn{1}{c}{Star with uniform density} &
\multicolumn{1}{c}{Star with polytropic fluid }&
\multicolumn{1}{c}{Galaxy rotation curves}
%\multicolumn{1}{c}{$\rho(0)$ (MeV$^4$) } &
%\multicolumn{1}{c}{$\sqrt{\theta}$($10^{18}$MeV$^{-1}$$\geqslant/\leqslant10^{-2}$)} 

\\
 
\cmidrule[0.4pt](r{0.125em}){1-1}
\cmidrule[0.4pt](lr{0.125em}){2-2}
\cmidrule[0.4pt](lr{0.125em}){3-3}
\cmidrule[0.4pt](lr{0.125em}){4-4}
\cmidrule[0.4pt](lr{0.125em}){5-5}
\cmidrule[0.4pt](lr{0.125em}){6-6}
\cmidrule[0.4pt](lr{0.125em}){7-7}
\cmidrule[0.4pt](lr{0.125em}){8-8}
\cmidrule[0.4pt](lr{0.125em}){9-9}
\cmidrule[0.4pt](lr{0.125em}){10-10}
\cmidrule[0.4pt](lr{0.125em}){11-11}
\cmidrule[0.4pt](lr{0.125em}){12-12}
$\sqrt\theta <\frac{1}{2}\sqrt{\frac{R}{GM}}R$ & $\kappa\geqslant10^2\sqrt{\theta} \;\; \rm and \;\;  \kappa\leqslant\sqrt{\theta}$ & 
$0.5$ kpc $< \sqrt\theta <  5$  kpc \\ 

 \bottomrule
\hline \hline
\end{tabular}
\end{center}
\caption{Compilation of the results shown through the paper for the constriction of NED parameter.}\label{tab:final}
\end{table*}
\end{center}
%%%%%%%%%%%%%%%%%%%%%%%%
%

The results provided by this paper, contributes to generate astrophysical bounds of the $\theta$ parameter which was studied previously in Ref. \cite{Rahaman:2010vv}, but leaving aside the contrast with real data. Of course, we remark that our results are not about NC, 
instead is just a model that was motivated by NC. NC it is a quantum-gravity theory which has not important effects at macroscopic scales. 

As a final note, we remark that the density profile inspired by NC can be used in cosmological analysis like structure formation, the statistics of the distribution of galaxy clusters, the temperature anisotropies of the cosmic microwave background radiation (CMB) and with other astrophysical studies. However, this work is in progress and will be reported elsewhere.

%%%%%%%%%%%%%%%%%%%%%%%%%%%%%%%%%%%%%%%%%%%%%%%
\begin{acknowledgments}
We would like to thank referee for his/her comments that were very useful to improve the manuscript. 
MARM thanks the helpful guide from Celia Escamilla-Rivera to improve the Mathematica notebooks used to make the calculations presented in Sec. \ref{GC}.
The authors acknowledge support from SNI-M\'exico, PIFI and PROMEP with number UAZ-CA-205.  JCL-D acknowledge support from F-PROMEP-39/Rev-03. MAG-A acknowledge support from CONACYT research fellow and Instituto Avanzado de Cosmolog\'ia (IAC) collaborations.
\end{acknowledgments}

\bibliography{librero1}

%%%%%%%%%%%%%%%%%%%%%%%%%%%%%%%
\appendix 
\section{PISO, NFW, Burkert and NED Rotation Curves} \label{Ap}
%%%%%%%%%%%%%%%%%%%%%%%%%%%%%%%

Here we give a brief summary of three important models found in the literature \cite{piso,NFW,Burkert} and the NED rotation curve model. 
In general, given a dark matter model with a mass profile, $M_{DM}(r)$, a rotation curve is described by \eqref{rotvel} written in the following form
\begin{equation}
V_{\rm DM}^{2}(r)= \frac{G M_{\rm DM}(r)}{r}, \label{velDM}
\end{equation}
where the mass at a given radius $r$ is given by
\begin{equation}
M_{\rm DM}(r)= 4\pi \int_0^r dr' r'^2 \rho_{DM}(r'/r_{c}) \label{massDM} .
\end{equation}
The function $\rho_{DM}(r/r_{c})$ is a function of $r/r_{c}$ multiplied by $\rho_{c}$.
Parameters $\rho_{c}$ and $r_{c}$ characterize the DM density profile. The density at the core of the halo is $\rho_c$ and $r_c$ is its scale length. It have turned out that the quantity 
\begin{equation}
\mu_{DM}(r)= \rho_{c}\, r_{c} \label{muDM}
\end{equation}
is almost a constant when we fit to observations.

\begin{enumerate}[(a)]

\item {\bf Pseudo isothermal profile.}
Here we consider that DM density profile is given by PISO \cite{piso} written as:
\begin{equation}
\rho_{\rm PISO}(r)=\frac{\rho_p}{1+(r/r_p)^{2}}. \label{PIP}
\end{equation}
With this density profile it is possible to obtain the mass at a given radius $r$
\begin{equation}
M_{\rm PISO}(r)= 4\pi r_{p}^{3}\rho_p \left(  \frac{r}{r_p}-\arctan\left(\frac{r}{r_p}\right)\right), \label{masspiso}
\end{equation}
and
from \eqref{velDM} together with \eqref{masspiso} we have
\begin{equation}
V_{\rm PISO}^{2}(r)= \frac{G M_{\rm PISO}(r)}{r}, \label{velpiso}
\end{equation}
known as PISO rotation velocity \cite{piso}.

\item {\bf Navarro-Frenk-White profile.}
Another interesting case (motivated by cosmological $N$-body simulations) is the NFW density profile \cite{NFW}, which is given by \cite{NFW}:
\begin{equation}
\rho_{\rm NFW}(r)=\frac{\rho_n}{(r/r_n)(1 + r/r_n)^{2}}. \label{NFW}
\end{equation}
From \eqref{massDM} with the above density profile we have
\begin{equation}
 M_{\rm NFW}(r)= 4\pi r_{n}^{3}\rho_n \left(  \ln(1+r/r_n) - \frac{r/r_n}{1+r/r_n} \right). \label{massNFW}
\end{equation}

From \eqref{velDM}, together with \eqref{massNFW} we obtain the following rotation velocity:
\begin{equation}
V_{\rm NFW}^{2}(r)=\frac{G M_{\rm NFW}(r)}{r}, \label{velnfw}
\end{equation}
this equation is known as NFW rotation velocity.

\item {\bf Burkert profile.}
Another density profile proposed by Burkert \cite{Burkert} is:
\begin{equation}
\rho_{\rm Burk}=\frac{\rho_{b}}{(1+r/r_b)(1+(r/r_b)^{2})}. \label{Burk}
\end{equation}
Again from \eqref{massDM} with the above density profile we have
\begin{eqnarray}
 M_{\rm Burk}(r) &=& 4\pi r_{b}^{3}\rho_b \frac{1}{4} \left( \ln\left[ (1+r/r_b)^2 (1+(r/r_b)^2)\right]  \right. \nonumber \\
 & & \left. - 2 \tan^{-1}(r/r_b) \right). \label{massBurk}
\end{eqnarray}
Again, from \eqref{velDM}, together with \eqref{Burk} we obtain the following rotation velocity:
\begin{equation}
V_{\rm Burk}^{2}(r) = \frac{G M_{\rm Burk}(r)}{r}, \label{velbur}
\end{equation}
which is known as Burkert rotation velocity \cite{Burkert}.

\item {\bf NED profile.}

Here we consider that DM density profile is given by NED density profile:
\begin{equation}
\rho_{\rm NED}(r)=\rho_{NED} \exp(-r^2/4r^2_{NED}), \label{rhoNED}
\end{equation}
where $r_{NED}=\sqrt \theta$.
As before with this density profile it is possible to obtain the mass at a given radius $r$
\begin{eqnarray}
M_{\rm NED}(r) &=& 4\pi r_{NED}^{3} \, \rho_{NED} 2 
\left[
\sqrt\pi \, \text{erf} ( r/2 r_{NED})  \right. \nonumber \\
 && \left. - ( r/r_{NED}) \exp (-r^2/4 r_{NED}^2)
\right], \label{massNED}
\end{eqnarray}
and
from \eqref{velDM} together with \eqref{massNED} we have
\begin{equation}
V_{\rm NED}^{2}(r)= \frac{G M_{\rm NED}(r)}{r}, \label{velNED}
\end{equation}
known as NED rotation velocity.

\item {\bf WaveDM profile.}

Here we consider that DM density profile is given by WaveDM density profile\cite{2017arXiv170100912B}:
\begin{equation}
\rho_{\rm WDM}(r)=\frac{\rho_w}{(1+(r/r_w)^2)^{8}}. \label{eq:WaveDM}
\end{equation}
As before with this density profile it is possible to obtain the mass at a given radius $r$
\begin{eqnarray}
M_{\rm WDM}(r) &=& 4\pi r_{w}^{3} \, \rho_{w}  
\left[
\frac{1}{215040 \left(\left(\frac{r}{{r_w}}\right)^2+1\right)^7} 
\right. \nonumber \\
&& \times \left(
48580 \left(\frac{r}{{r_w}}\right)^3-\frac{3465 r}{{r_w}}
\right. \nonumber \\
&&  + 101376 \left(\frac{r}{{r_w}}\right)^7+92323 \left(\frac{r}{{r_w}}\right)^5
\nonumber \\
&& + 23100 \left(\frac{r}{{r_w}}\right)^{11}+65373 \left(\frac{r}{{r_w}}\right)^9
\nonumber \\
&& + 3465 \left(\frac{r}{r_w}\right)^{13}
\nonumber \\
 && \left. 
+3465 \left(\left(\frac{r}{r_w}\right)^2+1\right)^7 \tan ^{-1}\left(\frac{r}{r_w}\right)
\right], 
\nonumber \\
\label{massWDM}
\end{eqnarray}
and
from \eqref{velDM} together with \eqref{massWDM} we have
\begin{equation}
V_{\rm WDM}^{2}(r)= \frac{G M_{\rm WDM}(r)}{r}, \label{velWDM}
\end{equation}
known as WaveDM rotation velocity.

\end{enumerate}

\end{document}